\documentclass[twocolumn]{aastex61}
\usepackage{comment}
\usepackage{amsmath,amssymb}
\usepackage{graphicx}
\usepackage{textcomp}
\usepackage{booktabs}
\usepackage{bm}

\newcommand\aastex{AAS\TeX}

\newcommand\ltsima{$\; \buildrel <\over\sim \;$}
\newcommand\simlt{\lower.5ex\hbox{\ltsima}}
\newcommand\gtsima{$\; \buildrel >\over\sim \;$}
\newcommand\simgt{\lower.5ex\hbox{\gtsima}}

\shorttitle{\aastex\ MOA-II FFP mass function}
\shortauthors{Sumi et al.}

\defcitealias{Koshimoto2023}{K23}
\defcitealias{Koshimoto2021}{Koshimoto+21a}
\defcitealias{Gould2022}{Gould+22}

\begin{document}

\title{Free-Floating planet Mass Function from MOA-II 9-year survey towards the Galactic Bulge}

\author[0000-0002-4035-5012]{Takahiro Sumi}
\affil{Department of Earth and Space Science, Graduate School of Science, Osaka University, Toyonaka, Osaka 560-0043, Japan. e-mail: {\tt sumi@ess.sci.osaka-u.ac.jp}}
\author[0000-0003-2302-9562]{Naoki koshimoto}
\affiliation{Department of Earth and Space Science, Graduate School of Science, Osaka University, Toyonaka, Osaka 560-0043, Japan}
\affiliation{Code 667, NASA Goddard Space Flight Center, Greenbelt, MD 20771, USA}
\affiliation{Department of Astronomy, University of Maryland, College Park, MD 20742, USA}
\author{David P.~Bennett}
\affiliation{Code 667, NASA Goddard Space Flight Center, Greenbelt, MD 20771, USA}
\affiliation{Department of Astronomy, University of Maryland, College Park, MD 20742, USA}
\author{Nicholas J. Rattenbury}
\affiliation{Department of Physics, University of Auckland, Private Bag 92019, Auckland, New Zealand}


\author{Fumio Abe}
\affiliation{Institute for Space-Earth Environmental Research, Nagoya University, Nagoya 464-8601, Japan}
\author{Richard Barry}
\affiliation{Code 667, NASA Goddard Space Flight Center, Greenbelt, MD 20771, USA}
\author{Aparna Bhattacharya}
\affiliation{Code 667, NASA Goddard Space Flight Center, Greenbelt, MD 20771, USA}
\affiliation{Department of Astronomy, University of Maryland, College Park, MD 20742, USA}
\author{Ian A. Bond}
\affiliation{Institute of Natural and Mathematical Sciences, Massey University, Auckland 0745, New Zealand}
\author{Hirosane Fujii}
\affiliation{Institute for Space-Earth Environmental Research, Nagoya University, Nagoya 464-8601, Japan}
\author{Akihiko Fukui}
\affiliation{Department of Earth and Planetary Science, Graduate School of Science, The University of Tokyo, 7-3-1 Hongo, Bunkyo-ku, Tokyo 113-0033, Japan}
\affiliation{Instituto de Astrof\'isica de Canarias, V\'ia L\'actea s/n, E-38205 La Laguna, Tenerife, Spain}
\author{Ryusei Hamada}
\affiliation{Department of Earth and Space Science, Graduate School of Science, Osaka University, Toyonaka, Osaka 560-0043, Japan}
\author{Yuki Hirao}
\affiliation{Institute of Astronomy, Graduate School of Science, The University of Tokyo, 2-21-1 Osawa, Mitaka, Tokyo 181-0015, Japan}
\author{Stela Ishitani Silva}
\affiliation{Department of Physics, The Catholic University of America, Washington, DC 20064, USA}
\affiliation{Code 667, NASA Goddard Space Flight Center, Greenbelt, MD 20771, USA}
\author{Yoshitaka Itow}
\affiliation{Institute for Space-Earth Environmental Research, Nagoya University, Nagoya 464-8601, Japan}
\author{Rintaro Kirikawa}
\affiliation{Department of Earth and Space Science, Graduate School of Science, Osaka University, Toyonaka, Osaka 560-0043, Japan}
\author{Iona Kondo}
\affiliation{Department of Earth and Space Science, Graduate School of Science, Osaka University, Toyonaka, Osaka 560-0043, Japan}
\author{Yutaka Matsubara}
\affiliation{Institute for Space-Earth Environmental Research, Nagoya University, Nagoya 464-8601, Japan}
\author{Shota Miyazaki}
\affiliation{Institute of Space and Astronautical Science, Japan Aerospace Exploration Agency, 3-1-1 Yoshinodai, Chuo, Sagamihara, Kanagawa 252-5210, Japan}
\author{Yasushi Muraki}
\affiliation{Institute for Space-Earth Environmental Research, Nagoya University, Nagoya 464-8601, Japan}
\author{Greg Olmschenk}
\affiliation{Code 667, NASA Goddard Space Flight Center, Greenbelt, MD 20771, USA}
\author{Cl\'ement Ranc}
\affiliation{Sorbonne Universit\'e, CNRS, UMR 7095, Institut d'Astrophysique de Paris, 98 bis bd Arago, 75014 Paris, France}
\author{Yuki Satoh}
\affiliation{Department of Earth and Space Science, Graduate School of Science, Osaka University, Toyonaka, Osaka 560-0043, Japan}
\author{Daisuke Suzuki}
\affiliation{Department of Earth and Space Science, Graduate School of Science, Osaka University, Toyonaka, Osaka 560-0043, Japan}
\author{Mio Tomoyoshi}
\affiliation{Department of Earth and Space Science, Graduate School of Science, Osaka University, Toyonaka, Osaka 560-0043, Japan}
\author{Paul . J. Tristram}
\affiliation{University of Canterbury Mt.\ John Observatory, P.O. Box 56, Lake Tekapo 8770, New Zealand}
\author{Aikaterini Vandorou}
\affiliation{Code 667, NASA Goddard Space Flight Center, Greenbelt, MD 20771, USA}
\affiliation{Department of Astronomy, University of Maryland, College Park, MD 20742, USA}
\author{Hibiki Yama}
\affiliation{Department of Earth and Space Science, Graduate School of Science, Osaka University, Toyonaka, Osaka 560-0043, Japan}
\author{Kansuke Yamashita}
\affiliation{Department of Earth and Space Science, Graduate School of Science, Osaka University, Toyonaka, Osaka 560-0043, Japan}
\collaboration{(MOA collaboration)}


\begin{abstract}
We present the first measurement of the mass function of free-floating planets (FFP) or very wide orbit planets down to
an Earth mass, from the MOA-II microlensing survey in 2006-2014.
Six events are likely to be due to planets with Einstein radius crossing times, 
$t_{\rm E}<0.5$days, and the shortest has $t_{\rm E} = 0.057\pm 0.016$days and
an angular Einstein radius of $\theta_{\rm E} = 0.90\pm 0.14\mu$as. 
We measure the detection efficiency depending on both $t_{\rm E}$ and $\theta_{\rm E}$ with image
level simulations for the first time.
These short events 
are well modeled by a power-law mass function, 
$dN_4/d\log M = (2.18^{+0.52}_{-1.40})\times (M/8\,M_\earth)^{-\alpha_4}$ dex$^{-1}$star$^{-1}$ with
 $\alpha_4 = 0.96^{+0.47}_{-0.27}$ for $M/M_\odot < 0.02$.
This implies a total of $f=  21^{+23}_{-13}$ FFP or very wide orbit planets 
of mass $0.33<M/M_\earth < 6660$ per star, with a total mass of
$80^{+73}_{-47} M_\earth$ per star.
The number of FFPs is $19_{-13}^{+23}$ times the number of planets 
in wide orbits (beyond the snow line), while the total masses are of the same order.
This suggests that the FFPs have been ejected from bound planetary systems that
may have had an initial mass function with a power-law index of $\alpha\sim 0.9$, which would imply
a total mass of $171_{-52}^{+80} M_\earth$ star$^{-1}$.
This model predicts that Roman Space Telescope will detect $988^{+1848}_{-566}$ FFPs with masses down to that of Mars (including
$575^{+1733}_{ -424}$  with $0.1 \le M/M_\earth \le 1$).
The \cite{sumi2011} large Jupiter-mass FFP population is excluded.
\end{abstract}




\section{Introduction} \label{sec:intro}

Gravitational microlensing observations toward the Galactic bulge (Galactic Bulge) enable exoplanet searches \citep{mao1991,gaudi-ogle109,bennett-ogle109, suzuki2016, kos21b}, 
and the measurement of the stellar and sub-stellar mass functions (MFs) \citep{pac91,sumi2011, Mroz17,Mroz19, Mroz20a}.

\cite{sumi2011}  first interpreted the detection of short
Einstein radius crossing time ($0.5<t_{\rm E}/{\rm day}<2$) 
microlensing events as evidence for the existence of a population of free-floating planets (FFP) and/or wide orbit planets.
While that analysis was limited by the small number of events found in a 2 year subset of the survey by the Microlensing Observation in Astrophysics (MOA) group \citep{sumi03} in collaboration 
with Optical Gravitational Lensing Experiment (OGLE) \citep{uda94},
it opened up the field of FFP studies using microlensing.

\cite{Mroz17} extended the work by using a larger sample from 5 years of the OGLE survey.
They discovered 6 events with timescales shorter ($t_{\rm E}\sim0.2$ day) than those in the previous work.
These events are separated from the longer events by a gap around $t_{\rm E}\sim0.5$ day 
which implying the possibility of a several Earth-mass FFP population.

These studies are based on distribution of $t_{\rm E}$, in which
$t_{\rm E}$ is proportional to the square root of the lens mass $M$ as follows,
\begin{equation}
  \label{eq:tE}
   \begin{split}
t_{\rm E}&=
\frac{ \sqrt{\kappa M \pi_{\rm rel}}}{\mu_{\rm rel}}\\
&= 0.1\, {\rm day}  \left(  \frac{M}{5 M_\earth} \right)^{1/2} 
 \left(  \frac{\pi_{\rm rel}}{18\mu \rm as} \right)^{1/2} 
 \left(  \frac{\mu_{\rm rel}}{5 \rm mas\,yr^{-1}} \right)^{-1}.
   \end{split}
   \end{equation}
Here, $\kappa =4G/(c^2 {\rm au})= 8.144 {\rm mas}/M_\odot$  and we expect $t_{\rm E}\sim0.1$ day assuming typical value of the lens-source relative parallax: 
$\pi_{\rm rel}=\pi_{\rm l}^{-1}- \pi_{\rm s}^{-1}=1$ au$(D_{\rm l}^{-1}-D_{\rm s}^{-1})=18\mu$as for the bulge lens and a typical value of the lens-source relative proper motion in the direction of the Galactic center of $\mu_{\rm rel}= 5$ mas\,yr$^{-1} $.
The lens mass $M$, the distance $D_{\rm l}$ to the lens and the relative proper motion $\mu_{\rm rel}$ are degenerate in the observable  $t_{\rm E}$. ($D_s$ is the distance to the source star.)
This means that the mass function of the lens population has to be determined
statistically, assuming a model of the star population density and velocities in the Galaxy.

\cite{Mroz18} found the first short ($t_{\rm E}=0.32$ day) event showing the Finite Source (FS) effect,
i.e., a finite source and a single point lens (FSPL), 
 in which one can measure a FS parameter $\rho=\theta_*/\theta_{\rm E}$.
 Here $\theta_*$ is the angler source radius which can be estimated 
 from an empirical relation with the source magnitude and color.
 The $\theta_{\rm E}$ is an angular Einstein radius given by
 \begin{equation}
  \label{eq:thetaE}
\theta_{\rm E} = \frac{\mu_{\rm rel}}{t_{\rm E}} = \sqrt{\kappa M \pi_{\rm rel}}.
\end{equation}

This value of $\theta_{\rm E}$ can give us an inferred mass of the lens with better accuracy as we can eliminate one of the three-fold degenerate terms
which affect $t_{\rm E}$, namely, $\mu_{\rm rel}$:
 \begin{equation}
  \label{eq:M_thetaE_pirel}
M=\frac{\theta_{\rm E}^2}{\kappa \pi_{\rm rel}}  =  
5 M_\earth  \left(  \frac{\theta_{\rm E}} {1.5\mu \rm as} \right)^2 
\left(  \frac{\pi_{\rm rel}}{18\mu \rm as} \right)^{-1}.
\end{equation}
While the inclusion of the angular Einstein radius, $\theta_E$, enables tighter constraints
on the lens masses, it adds a complication to a statistical analysis of FFP properties
because the microlensing event detection efficiency depends on both $t_E$ and 
$\theta_E$ (or equivalently $t_E$ and $\rho$).

So far, six short FSPL events have been discovered 
\citep{Mroz18,Mroz19b,Mroz20b,Mroz20c,Kim2021,Ryu2021}.
All of these have $\theta_{\rm E}<10\,\mu$as, implying that their lenses
 are most likely of planetary mass.
All of these sources are red giants with the exception of the sub-giant source for OGLE-2016-BLG-1928
because their angular radii, i.e., cross-section,  
are significantly larger than main sequence stars.

\cite{Mroz20b} found the short FSPL event, OGLE-2016-BLG-1928, 
with the smallest value of $\theta_{\rm E}=  0.842 \pm 0.064 \mu$as to date. 
Its lens is the first terrestrial mass FFP candidate and
the first evidence of such a population.

\cite{Kim2021} began a new approach to probing the FFP population 
by focusing on analyzing the $\theta_{\rm E}$ distribution in events
with giant sources.
\cite{Ryu2021} found a gap at $10<\theta_{\rm E}/\mu {\rm as} <30$ 
in the cumulative $\theta_{\rm E}$ distribution, which
suggests a separation between the planetary mass population and
 other known populations, like brown dwarfs.

\cite{Gould2022} completed the analysis of 29 FSPL giant-source events 
found in the 2016-2019 KMTNet survey. 
 They presented the $\theta_{\rm E}$  distributions
down to  $\theta_{\rm E}=4.35\,\mu$as and  confirmed that there is a clear gap 
in the distribution of $\theta_{\rm E}$ at $9<\theta_{\rm E}/\mu{\rm as}<26$. 
They note that it is consistent with the gap in the $t_{\rm E}$ distribution
shown by \cite{Mroz17}, indicating 
the existence of the low mass FFP population.
They 
used what they refer to as a ``relative detection efficiency" that depends only on
$\theta_{\rm E}$, but not $t_{\rm E}$, to model the $\theta_{\rm E}$ distribution with a power law MF
for the FFP and found 
$dN_{\rm FFP}/d\log M=(0.4\pm 0.2)(M/38M_\earth)^{-p}$ dex$^{-1}$\,star$^{-1}$, 
using a power law with $0.9\lesssim p\lesssim1.2$. This range of the 
power, $p$, was estimated based on consideration of
possible formation mechanisms, rather than a measurement.
This would imply that the number of FFPs is at least 
an order of magnitude larger than the number of known bound planets.

We note that the \cite{Gould2022} result cannot be considered a measurement for
a the following reasons. First, the true detection efficiency depends on both $t_{\rm E}$ and $\theta_{\rm E}$,
and it is difficult to see how any selection criteria could remove the $t_{\rm E}$ dependence.
As we discuss below in Section~\ref{sec:L_longtE} and in \citetalias{Koshimoto2023} 
one can integrate over the $t_{\rm E}$ dependence of the detection efficiency to
obtain an integrated detection efficiency. However, the integration over short $t_{\rm E}$ values
depends on the FFP mass function. However, \cite{Gould2022} seem to avoid this
difficulty by simply adopting an analytic formula for the ``relative detection efficiency"
depending only on $\theta_{\rm E}$. The \cite{Gould2022} paper gives no justification
for this analytic formula.


%



In this paper, we present the distributions  $\theta_{\rm E}$ and $t_{\rm E}$
values for the microlensing events 
toward the Galactic Bulge from 9 years of the MOA-II survey.
We also present the first measurement of MF of the planetary mass objects 
using the $t_{\rm E}$ distribution.
We describe the data in section \S\,\ref{sec:Data}.
We show the $\theta_{\rm E}$ distribution
in section \S\,\ref{sec:FSPL}.
We present the $t_{\rm E}$ distribution and the best-fit MF in \S\,\ref{sec:timescale}. 
The discussion and conclusions are given in section \S\,\ref{sec:discussionAndSummary},
and we compare the integrated detection efficiency in Appendix~\ref{sec-append}.


\section{Data} \label{sec:Data}
We use the microlensing sample selected from the MOA-II high cadence photometric survey 
toward the Galactic Bulge in the 2006-2014 seasons \citep[][hereafter \citetalias{Koshimoto2023}]{Koshimoto2023}.
MOA-II uses the 1.8-m MOA-II telescope which has 
a 2.18 deg$^2$ field of view (FOV) and which is
located at the Mt.\ John University
Observatory, New Zealand\footnote{\url{https://www.massey.ac.nz/~iabond/moa/alerts/}}.


\citetalias{Koshimoto2023} used an analysis method similar to what was used by \cite{sumi2011,sumi2013},
but includes a correction of systematic errors and takes into 
account the finite source effect.
They applied a de-trending code to all light curves to remove the systematic errors that correlate with seeing and airmass due to
differential refraction, differential extinction and relative proper motion of stars
in the same way as in \cite{bennett2012} and \cite{sumi2016}. 
These corrections are important as they result in higher confidence
in the light curve fitting parameters.


 \citetalias{Koshimoto2023} selected light curves with a single instantaneous 
brightening episode and a flat constant baseline, which can be well 
fit with a point-source point-lens (PSPL) microlensing model \citep{pac86}.
In addition to PSPL, they modeled the events with a FSPL model \citep{Bozza2018}, which is especially important for short events.
These are the major improvements compared to the previous analysis in \cite{sumi2011,sumi2013}
in addition to the extension of the survey duration.

Although they identified 6,111 microlensing candidates, 
they selected only 3,554 and 3,535 objects as the statistical sample using the two relatively strict criteria CR1 and CR2, 
respectively.
Here, CR2 was defined as the stricter criteria compared to 
their nominal criteria CR1 to check the effect of the choice of 
the criteria on a statistical study.
These strict criteria ensure that $t_{\rm E}$ is well constrained
for each event and reject any contamination. 

\cite{sumi2011} reported 10 short events with $t_{\rm E}<2$ days in the 2006-2007 dataset.
Only 5 and 4 events survived following the application of CR1 and CR2, respectively. 
This is because the fitting results changed due to the re-reduction of the dataset.
On the other hand, two events are newly found resulting 7 and 6 events following the application of CR1 and CR2, respectively.
As a result, the excess at $t_{\rm E}=0.5-2$ day in the $t_{\rm E}$ distribution is not significant anymore, however, an even shorter event MOA-9y-6057 ($t_{\rm E}=0.22 \pm 0.06$ day) is added.


\begin{deluxetable*}{lrrrrrc}
\tabletypesize{\scriptsize}
\tablecaption{Comparison of parameters of short FS events with known FFP candidates.
\label{tbl:candlistFFP}
}
\tablewidth{0pt}
\tablehead{
\colhead{field-chip-sub-ID} &
\colhead{$t_{\rm E}$} & 
\colhead{$\rho$} &
\colhead{$I_{\rm s,0}$} &
\colhead{$\theta_*$} &
\colhead{$\theta_{\rm E}$} & 
\colhead{reference} \\
\colhead{} & 
\colhead{$(\rm day)$} & 
\colhead{} &
\colhead{${(\rm mag)}$} & 
\colhead{($\mu$as)} & 
\colhead{($\mu$as)} &
\colhead{} 
}
\startdata
MOA-9y-5919     &       0.057 $\pm$ 0.016   &  1.40 $\pm$ 0.46   &      17.23   &   1.26 $\pm$   0.48 &  0.90 $\pm$   0.14 & \citetalias{Koshimoto2023} \\ 
MOA-9y-770      &       0.315 $\pm$ 0.017    &  1.08 $\pm$ 0.07    &      14.71   &   5.13 $\pm$   0.86  & 4.73 $\pm$   0.75 & \citetalias{Koshimoto2023} \\ 
\hline
OGLE-2016-BLG-1928 & 0.0288 $_{-0.0016}^{+0.0024}$  & 3.39$_{-0.11}^{+0.10}$  & 15.78 &  2.85 $\pm$ 0.20   & 0.842 $\pm$ 0.064  & \cite{Mroz20b} \\
KMT-2019-BLG-2073    & 0.272 $\pm$ 0.007      & 1.138 $\pm$ 0.012          & 14.45    &     5.43 $\pm$ 0.17   &  4.77 $\pm$ 0.19   & \cite{Kim2021} \\
KMT-2017-BLG-2820    & 0.288 $\pm$ 0.015    & 1.096 $\pm$ 0.079           & 14.31     &     7.05 $\pm$ 0.44 &  5.94 $\pm$ 0.37  & \cite{Ryu2021}\\
OGLE-2012-BLG-1323  & 0.155 $\pm$ 0.005     & 5.03 $\pm$ 0.07               &    14.09   &    11.9 $\pm$ 0.5      & 2.37 $\pm$ 0.10    & \cite{Mroz19b}\\
OGLE-2016-BLG-1540  &  0.320 $\pm$ 0.003    &  $1.65 \pm 0.01$             &    13.51   &   15.1 $\pm$ 0.8     &  9.2 $\pm$ 0.5     &   \cite{Mroz18}\\
OGLE-2019-BLG-0551 & 0.381 $\pm$ 0.017     & 4.49 $\pm$ 0.15              &  12.61    &    19.5 $\pm$ 1.6     &  4.35 $\pm$ 0.34   & \cite{Mroz20c} \\
\hline
MOA-9y-1944\tablenotemark{$a$}     &       1.594 $\pm$ 0.136   &  0.00928 $\pm$ 0.00032   &      20.14   &   0.43 $\pm$   0.10 &  46.1 $\pm$   10.5 & \citetalias{Koshimoto2023} \\ %
OGLE-2017-BLG-0560\tablenotemark{$a$}  & 0.905 $\pm$ 0.005     & 0.901 $\pm$ 0.005           &    12.47   &    34.9 $\pm$ 1.5      & 38.7 $\pm$ 1.6      & \cite{Mroz19b}\\
 \enddata
 \tablenotetext{a}{Likely Brown dwarf lens.}
\end{deluxetable*}

\begin{figure}
\begin{center}
\includegraphics[scale=0.35,keepaspectratio]{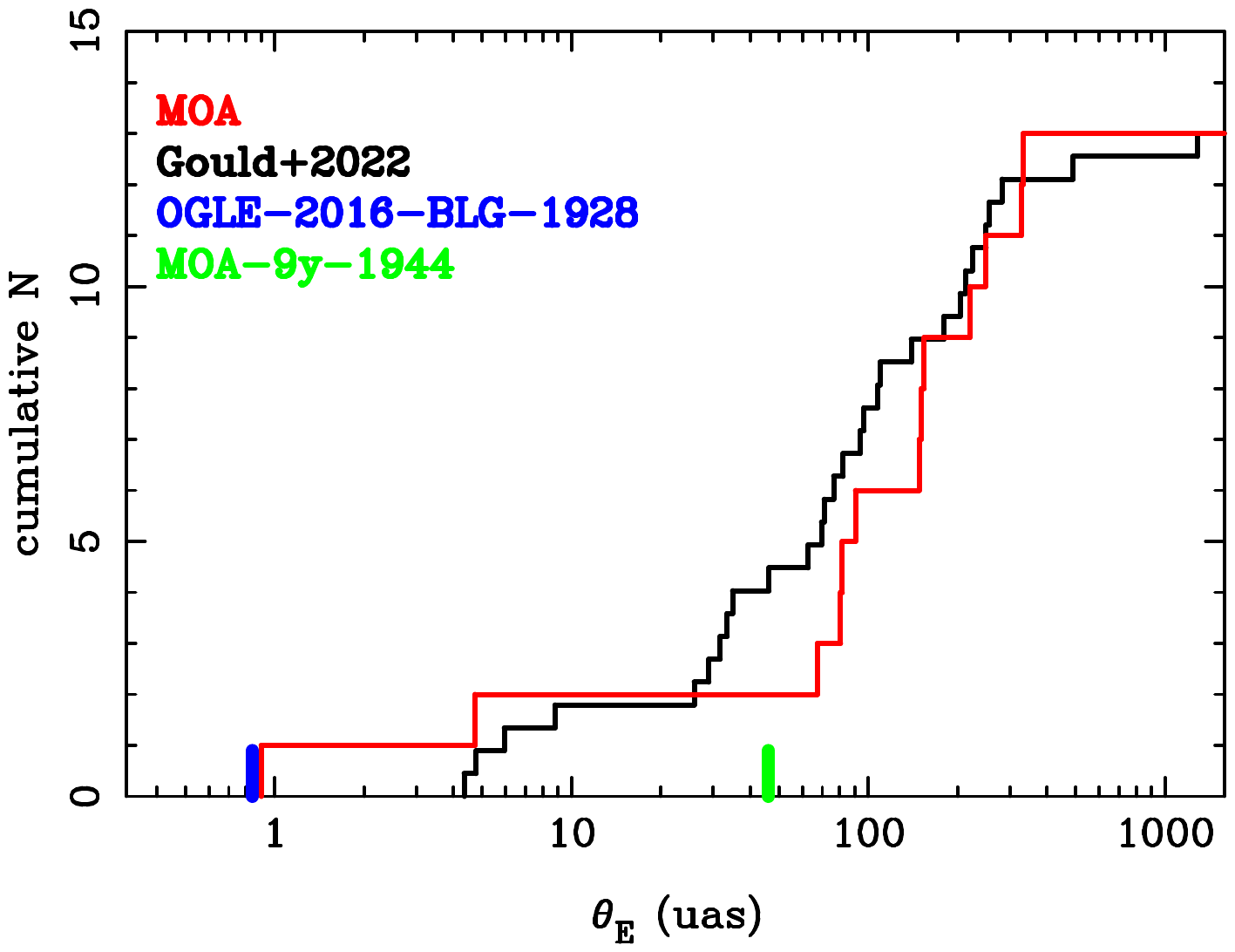}
\caption{
  \label{fig:thetaE}
Observed cumulative distribution of $\theta_{\rm E}$ for 13 FSPL events from MOA (red line)
and 29 FSPL events from KMTNet (black line) \citep{Gould2022}. The blue line indicates $\theta_{\rm E}=  0.842 \pm 0.064$ of 
terrestrial mass FFP candidate, OGLE-2016-BLG-1928 \citep{Mroz20b}.
}
\end{center}
\end{figure}

\begin{figure}
\begin{center}
\includegraphics[angle=0,width=8.5cm,keepaspectratio]{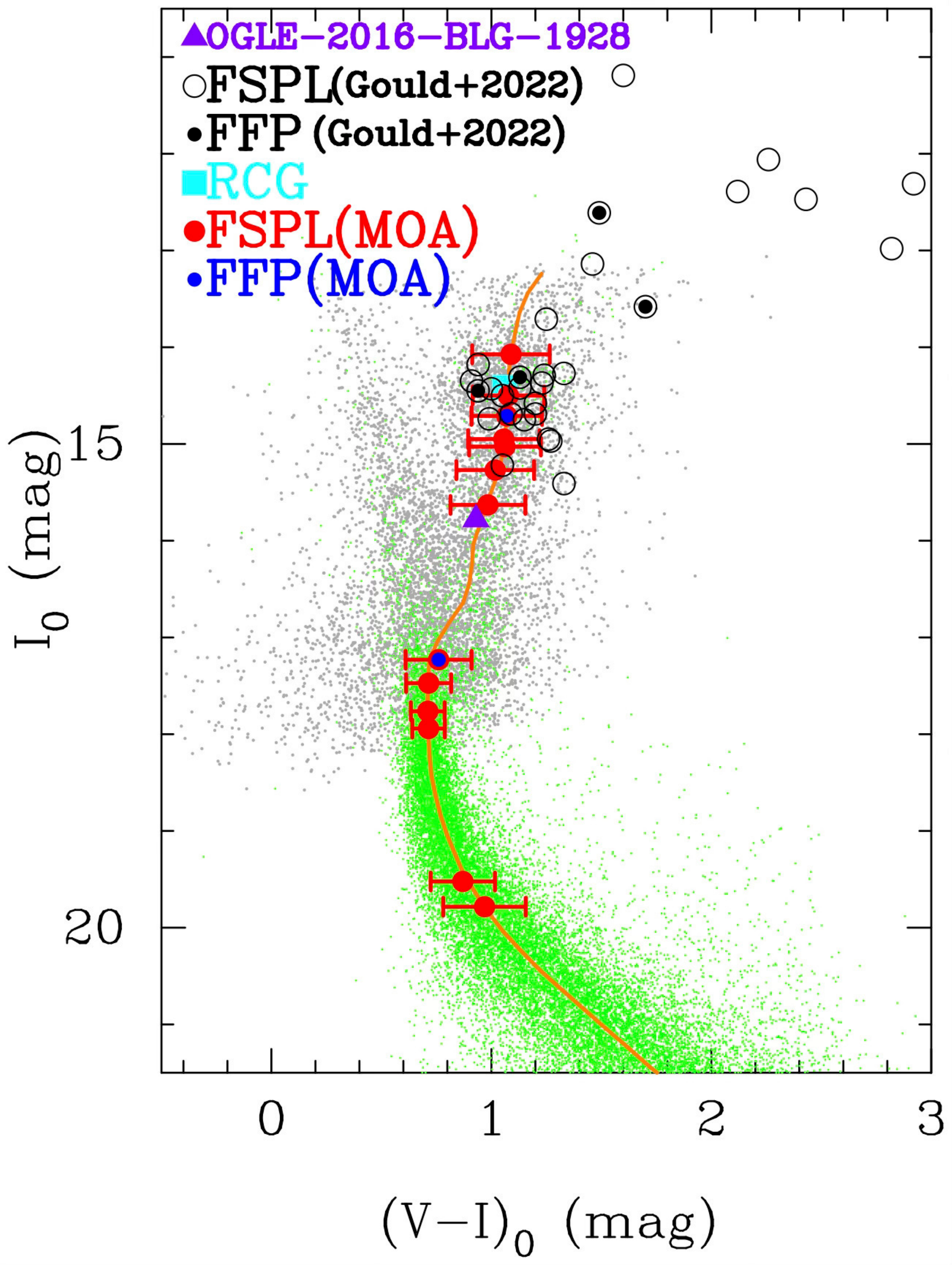}
\caption{
  \label{fig:CMDLDallGould}
Extinction free CMD of gb3-7-6. The orange curve is the isochrone matched to this subfield.
The cyan square is the RCG centroid. The red circles with error bars are sources of the 13 FSPL events in this work.
The blue filled circles indicates the 2 FFP candidates in this work.
The black open and filled circles are FSPL events and FFP events from \cite{Gould2022}, respectively.
The purple triangle indicates the source of terrestrial FFP,  OGLE-2016-BLG-1928S \citep{Mroz20b}.
}
\end{center}
\end{figure}

\section{Angular Einstein radius distribution}
\label{sec:FSPL}
There are 13 FSPL events with $\theta_{\rm E}$ measurements in the sample, including two FFP candidates, MOA-9y-5919 and MOA-9y-770, that have terrestrial and Neptune masses, respectively.
See \citetalias{Koshimoto2023} for the light curves and detailed parameters of the 13 events.

The red line in Figure \ref{fig:thetaE} indicates the cumulative distribution of 
$\theta_{\rm E}$ from Table 7 of \citetalias{Koshimoto2023}.
The black line indicates the distribution of 29 FFPs by \cite{Gould2022} normalized to 13 events as a comparison.
Although these can not be directly compared because these are not corrected for detection efficiencies, the general trends seen Figure \ref{fig:thetaE} 
may give us some insights. 

The distributions are consistent for $\theta_{\rm E}>30$\,$\mu$as, where the effect of the detection efficiencies are likely small.
There is a gap around $5<\theta_{\rm E}/{\rm \mu as}<70$ which is roughly consistent with 
the gap at  $10<\theta_{\rm E}/{\rm \mu as}<30$ found by  \cite{Ryu2021} and \cite{Gould2022}.
This gap confirmed the existence of the planetary mass population 
as distinct and separated from the stellar/brown dwarf population as 
indicated by \cite{Gould2022}.

The MOA cumulative distribution shows fewer events over $30<\theta_{\rm E}/{\rm \mu as}<70$ compared to \cite{Gould2022}.
This may be just due to the small number of statistics.
But note that \citetalias{Koshimoto2023} found a brown dwarf candidate MOA-9y-1944 with $\theta_{\rm E}= 46.1 \pm 10.5\,{\rm \mu as}$ although this is not
in the final sample for statistical analysis because the source magnitude of $I_{\rm s}=21.91$ mag is fainter than the threshold of $I_{\rm s}<21.4$ mag.

In our sample, there is one event with a very small value of 
$\theta_{\rm E}$ of $0.90 \pm 0.14$\,$\mu$as.
This confirms the existence of the terrestrial mass population 
which gives rise to events such as 
OGLE-2016-BLG-1928 which has $\theta_{\rm E}=  0.842 \pm 0.064$ \citep{Mroz20b}.
These values are significantly smaller than the lower edge of $\theta_{\rm E}\sim  4.35$\,$\mu$as as reported in  \cite{Gould2022}.
This is partly a result of selection bias given that \cite{Gould2022} focused on the sample with super-giant sources,
see Figure \ref{fig:CMDLDallGould}.

We compare the parameters of these events to six known FFP candidates with  $\theta_{\rm E}$ measurements in Table \ref{tbl:candlistFFP}.
The sources of all known FFP candidates except OGLE-2016-BLG-1928 are 
red clump giants (RCGs) or red super-giants which have large $\theta_{\rm *}=5.4, 7.1, 11.9, 15.1, 19.5$ and $34.9$\,$\mu$as.
The magnification tend to be suppressed by large $\theta_{\rm *}$ with small $\theta_{\rm E}$, i.e., large $\rho$ as
$A_{\rm FS,max}= \sqrt{1+4/\rho^2}$ $(\rho>1)$ \citep{Maeder1973,Agol2003,Riffeser2006}.
For example, in case of the terrestrial mass lens 
with $\theta_{\rm E}\sim 1\mu$as, 
the maximum magnification will be only $A_{\rm FS, max}= 1.066, 1.039, 1.014, 1.009, 1.005$ and $1.002$ for the above values of $\theta_{\rm *}$, respectively.
Note that the source of the terrestrial FFP candidate event, 
OGLE-2016-BLG-1928S is a sub-giant with  $\theta_{\rm *}=2.37$\,$\mu$as.
It is important to search for short FSPL with sub-giants and 
dwarf sources to find low mass FFP.
There is no FSPL event with a red super-giant source 
in our sample because these are saturated in MOA image data.



\begin{figure}
\begin{center}
\includegraphics[scale=0.43,keepaspectratio]{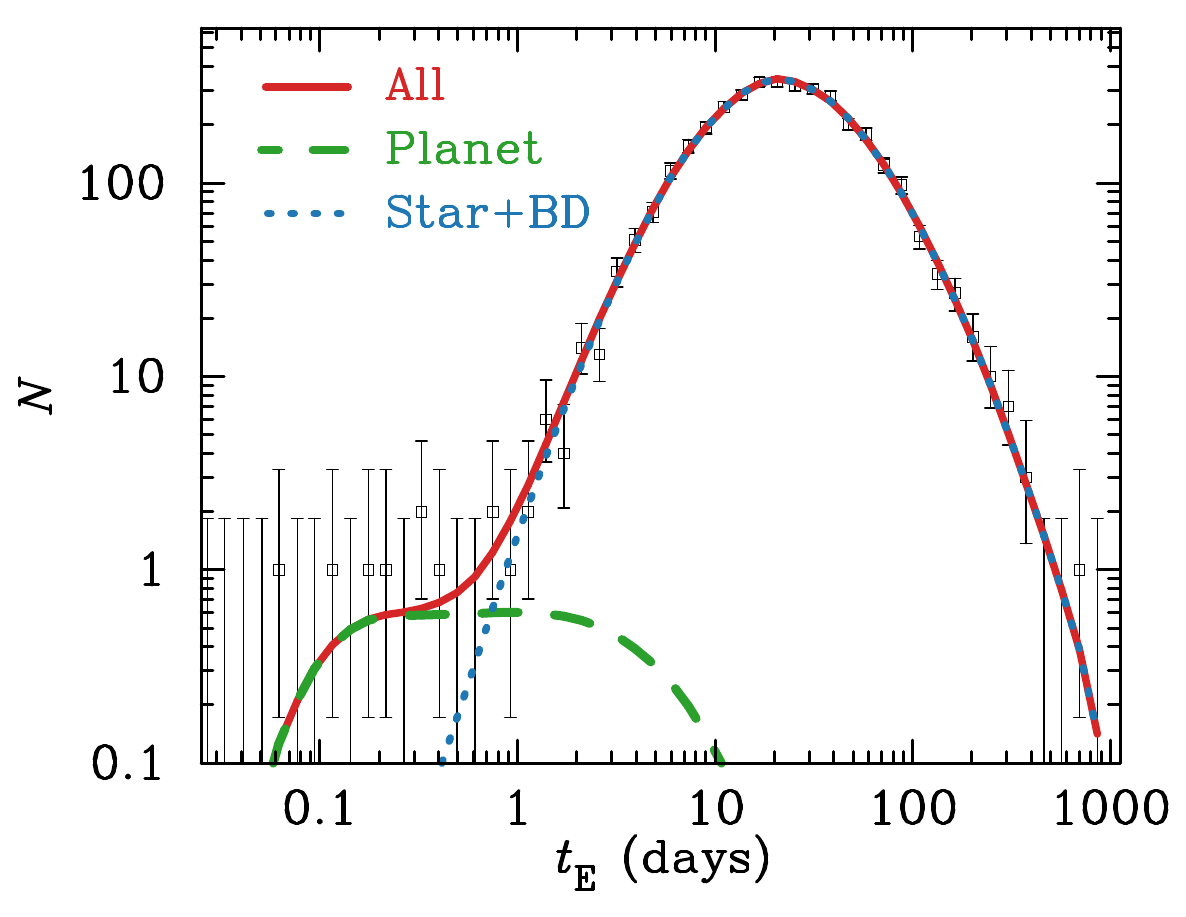}
\caption{
  \label{fig:tE}
The observed timescale $t_{\rm E}$ distribution passing criteria CR2 from the 9 year MOA-II survey.
The 1-$\sigma$ error bars and upper limits are based on the poisson distribution.
The red line indicates the best fit single lens model for all population. The blue dotted line represents the known
populations of stars, brown dwarfs, and stellar remnants, and the 
green dashed line represents the planetary mass population.
}
\end{center}
\end{figure}

\section{Likelihood analysis of mass function}
\label{sec:timescale}
In the final sample of \citetalias{Koshimoto2023}, there are 10 (12) short timescale events with $t_{\rm E} <1$~day after applying CR2 (CR1).
Figure \ref{fig:tE} shows the $t_{\rm E}$ distribution of the CR2 sample.
The distribution is roughly symmetric in $\log t_{\rm E}$, 
with a tail at $t_{\rm E}<0.5$.
This confirmed the existence of such short timescale events 
with $t_{\rm E}<0.5$ day as reported by \cite{Mroz17}.
In this section, we 
perform a likelihood analysis on each of the 3554 (CR1) and 3535 (CR2) events using a Galaxy model to 
constrain the mass function of lens objects.

We define the likelihood, ${\cal L}$, in Section \ref{sec:efficiency}. In Sections \ref{sec:MFknown} and \ref{sec:MFFFP}, we determine the mass function without and with 
 a planetary mass population, respectively, by minimizing $\chi^2 \equiv -2 \ln {\cal L}$. Although the absolute value of $\chi^2$ is not meaningful due to its dependence on an arbitrary normalization associated with our likelihood calculation, the fitting procedure is still statistically valid as the relative likelihood between two models, represented by $\Delta \chi^2$, is independent of the normalization.

Note that results of the likelihood analysis for sample CR1 and CR2 are very similar.
In the following sections, we show only the results for CR2 as our final results 
except in the tables.

\subsection{Likelihood}
\label{sec:efficiency}
Although our sample contains more than 3500 events, the mass function of planetary-mass objects is largely determined by the events with $t_{\rm E} < 1$ day, which account for about 0.3\% of these events.
We define two likelihoods: ${\cal L}_{\rm short}$ for short timescale events with the 
best-fit $t_{\rm E} < 1$ day, and ${\cal L}_{\rm long}$ for events with the best-fit $t_{\rm E} \geq 1~{\rm day}$. 
In our likelihood analysis, we use the combined likelihood ${\cal L} = {\cal L}_{\rm short} {\cal L}_{\rm long}$.

For ${\cal L}_{\rm long}$, we simply use the best-fit $t_{\rm E}$ values provided by \citetalias{Koshimoto2023}, which is similar to the approach by previous studies  \citep{sumi2011, Mroz17}.
This is because of (i) the relatively small uncertainties in $t_{\rm E}$, (ii) the effect of individual $t_{\rm E}$ uncertainties is statistically marginalized by the large number of events, (iii) the limited sensitivity to $\theta_{\rm E}$, and (iv) the minimal impact on our primary goal of measuring the mass function of planetary mass objects.

On the other hand, the situation is the opposite for the short events, ${\cal L}_{\rm short}$. That is:  (i)
the $t_{\rm E}$ uncertainties are relatively large due to their shorter magnification period
but they must be smaller than the event selection threshold listed in Table 2 of \citetalias{Koshimoto2023},
(ii) the number of events is very limited (12 for CR1 and 10 for CR2), and $t_{\rm E} <$ 1 day range is only sparsely covered in Figure \ref{fig:tE}. Thus, the number of $t_{\rm E} <$ 1 day events may not be sufficient to statistically marginalize the effect of $t_{\rm E}$ uncertainties of individual events in the likelihood analysis,
(iii) because the $\rho=\theta_*/\theta_{\rm E}$ values are generally much larger than those of longer timescale events, one may get beneficial constraints on $\theta_{\rm E}$ even when the $\theta_{\rm E}$ values are not well determined, and (iv) they play a crucial role in determining the mass function of planetary mass objects.
Therefore, we must use the joint probability distribution of $(t_{\rm E}, \theta_{\rm E})$ for each event derived by \citetalias{Koshimoto2023} using the Markov Chain Monte Carlo (MCMC) method for 
${\cal L}_{\rm short}$. 
However, the $(t_{\rm E}, \theta_{\rm E})$ probability distributions  for each event depends on the FFP 
mass function that we are trying to measure, while
the event detection efficiency also depends on both $t_{\rm E}$ and $\theta_{\rm E}$. 
So, the probability 
distribution for the $t_{\rm E}$ and $\theta_{\rm E}$ values for each event depends upon both the light curve data and the FFP mass function. Rather than running our light curve model MCMC calculations for the short events separately for 
every mass function model we consider, we simplify our calculations by using the `importance sampling" method of
Monte Carlo integration \citep{num_recipies}. 
This means that we run the
MCMC light curve models with weighting of the $\log t_{\rm E}$ and $\log \theta_{\rm E}$ distributions
given by an uninformative (and incorrect) ``prior\rlap," $p_0 (\log t_{\rm E}, \log \theta_{\rm E})$ 
that is uniform in both $\log t_{\rm E}$ and $\log \theta_{\rm E}$. 
A function like $p_0 (\log t_{\rm E}, \log \theta_{\rm E})$ 
is sometimes called an ``interim prior" \citep{foreman-mac14}, but we have not used it as a Bayesian prior.
Instead, we replace $p_0 (\log t_{\rm E}, \log \theta_{\rm E})$ with the
correct distribution over $\log t_{\rm E}$ and $\log \theta_{\rm E}$ for each mass function model in our
FFP mass function likelihood calculation.
The only Bayesian prior assumptions assumed in this analysis are the Galactic model assumptions discussed in
Section \ref{sec:MFknown} and the mass function model priors discussed in Section \ref{sec:MFFFP} .

We describe the simpler likelihood function for the long duration events, ${\cal L}_{\rm long}$, in Section \ref{sec:L_longtE}, and then we describe ${\cal L}_{\rm short}$ in Section \ref{sec:L_shorttE}.

\subsubsection{Likelihood for events with $t_{\rm E} \geq 1$ day} \label{sec:L_longtE}

We define the likelihood for events with $t_{\rm E} \geq 1$ day by
\begin{align}
    {\cal L}_{\rm long} \propto \prod_{i = 1}^{N_{\rm long}} {\cal G} (t_{{\rm E}, i} ; \Gamma), \label{eq:Llong}
\end{align}
where $i$ runs over all the $N_{\rm long}$ events that have the best-fit $t_{\rm E} \geq 1$ day in our sample ($N_{\rm long} = 3542$  for CR1 and $N_{\rm long} = 3525$ for CR2), 
and $t_{{\rm E}, i}$ is the best-fit $t_{\rm E}$ value for $i$th event given by \citetalias{Koshimoto2023}.

The function ${\cal G} (t_{\rm E} ; \Gamma)$ is the model's detectable event rate as a function 
of $t_{\rm E}$ with given model event rate $\Gamma$, combined for the 20 survey fields, given by
\begin{align}
    {\cal G} (t_{\rm E} ; \Gamma) = \sum_j w_j \, g_j (t_{\rm E} ; \Gamma_j) \label{eq:G_tE}.
\end{align}
Here, $j$ takes field index values
gb1 to gb21, except for gb6. See Table 1 of \citetalias{Koshimoto2023} for the location and properties of each field.
The weight $w_j$ for the $j$th field is given by 
\begin{align}
    w_j = \sum_{k \in j} n_{{\rm RC}, k}^2 f_{{\rm LF}, k} \label{eq:w_j},
\end{align}
where $k$ indicates a 1024 pixel $\times$ 1024 pixel subframe in the $j$th field ($k = 1, 2, ..., 80$), $n_{{\rm RC}, k}$ is the number density of RCGs in the $k$th subfield, $f_{{\rm LF}, k}$ is the fraction of stars with magnitude $I < 21.4$\,mag in the $k$th subfield, and $w_j$ is thus proportional to the expected event rate in the $j$th field. To calculate $f_{{\rm LF}, k}$,  
we used a combined luminosity function that uses the OGLE-III photometry map \citep{uda11} for bright stars and the {\it Hubble Space Telescope} data by \citep{hol98} for faint stars.

The function $g_j$ is the model's detectable event rate as a function of $t_{\rm E}$ for field $j$ as given by
\begin{align}
    g_j (t_{\rm E} ; \Gamma_j) &= \tilde{\epsilon}_j (t_{\rm E} ; \Gamma_j) \, \Gamma_j (t_{\rm E}), \label{eq:f_tE}
\end{align}
where $\tilde{\epsilon} (t_{\rm E} ; \Gamma)$ is the integrated detection efficiency of the survey as a function of $t_{\rm E}$. \citetalias{Koshimoto2023} demonstrated that when finite source effects are important, the detection efficiency,
$\epsilon (t_{\rm E}, \theta_{\rm E})$ is a function of two variables, $t_{\rm E}$ and $\theta_{\rm E}$.
Therefore, we must integrate over $\theta_{\rm E}$ to obtain the integrated detection efficiency, 
$\tilde{\epsilon}(t_{\rm E} ; \Gamma)$, which now depends upon the event rate and the mass function of the lens objects.
This gives
\begin{equation}
\tilde{\epsilon}_j (t_{\rm E} ; \Gamma) =  \int_{\theta_{\rm E}} \epsilon_{j} (t_{\rm E}, \theta_{\rm E}) \, \Gamma_j (\theta_{\rm E} | t_{\rm E})   d\theta_{\rm E}  \label{eq:eps_tE},
\end{equation}
where $\epsilon_{j} (t_{\rm E}, \theta_{\rm E})$ is the detection efficiency for events with $t_{\rm E}$ and $\theta_{\rm E}$ for $i$th field.
We use the detection efficiency $\epsilon_{j} (t_{\rm E}, \theta_{\rm E})$ estimated by the image level simulations in \citetalias{Koshimoto2023} for the 20 fields of the MOA-II 9-yr survey.

We consider the model event rate as functions of $t_{\rm E}$ and $(t_{\rm E},\theta_{\rm E})$, denoted by $\Gamma (t_{\rm E})$ and $\Gamma (t_{\rm E},\theta_{\rm E})$, respectively.
These are normalized functions so that their integrations give one, i.e., these are probability density functions of $t_{\rm E}$ and $(t_{\rm E},\theta_{\rm E})$, respectively. $\Gamma (\theta_{\rm E} | t_{\rm E}) = \Gamma (t_{\rm E}, \theta_{\rm E})/\Gamma (t_{\rm E})$ is the probability density of events with $\theta_{\rm E}$ given $t_{\rm E}$.
Thus, the calculation of $\tilde{\epsilon}(t_{\rm E} ; \Gamma)$ in Eq. (\ref{eq:eps_tE}) has to be done for every proposed MF
during the fitting procedure because $\Gamma (\theta_{\rm E} | t_{\rm E})$ depends on the MF.

The function $\Gamma_j (t_{\rm E}, \theta_{\rm E})$ for $j$th field can be separated from the MF \citep{han96},
\begin{align}
 \Gamma_j (t_{\rm E}, \theta_{\rm E}) = \int \gamma_j (t_{\rm E} M^{-1/2}, \theta_{\rm E} M^{-1/2}) \Phi (M) \sqrt{M} dM, \label{eq:G_tEthE}
\end{align}
where $\gamma_j (t_{\rm E}, \theta_{\rm E})$ is the event rate for lenses with
mass $1\,M_\sun$ and $\Phi (M)$ is the present-day MF (expressed as $dN/dM$).
Although substituting Eqs. (\ref{eq:eps_tE}) and (\ref{eq:G_tEthE}) makes the calculation of $g_j (t_{\rm E} ; \Gamma_j)$ in Eq. (\ref{eq:f_tE}) a double integral over $M$ and $\theta_{\rm E}$, \citetalias{Koshimoto2023} showed that the integration over $\theta_{\rm E}$ is largely avoidable during a fitting procedure by switching the order of the integrals and calculating the integral over $\theta_{\rm E}$ before the fitting.

We calculate $\gamma_j (t_{\rm E}, \theta_{\rm E})$ for each field using the density and velocity distribution of stars from the latest parametric Galactic model toward 
the Galactic Bulge based on Gaia and microlensing data \citep{Koshimoto2021}.

\begin{figure}
\begin{center}
\includegraphics[width=8.5cm]{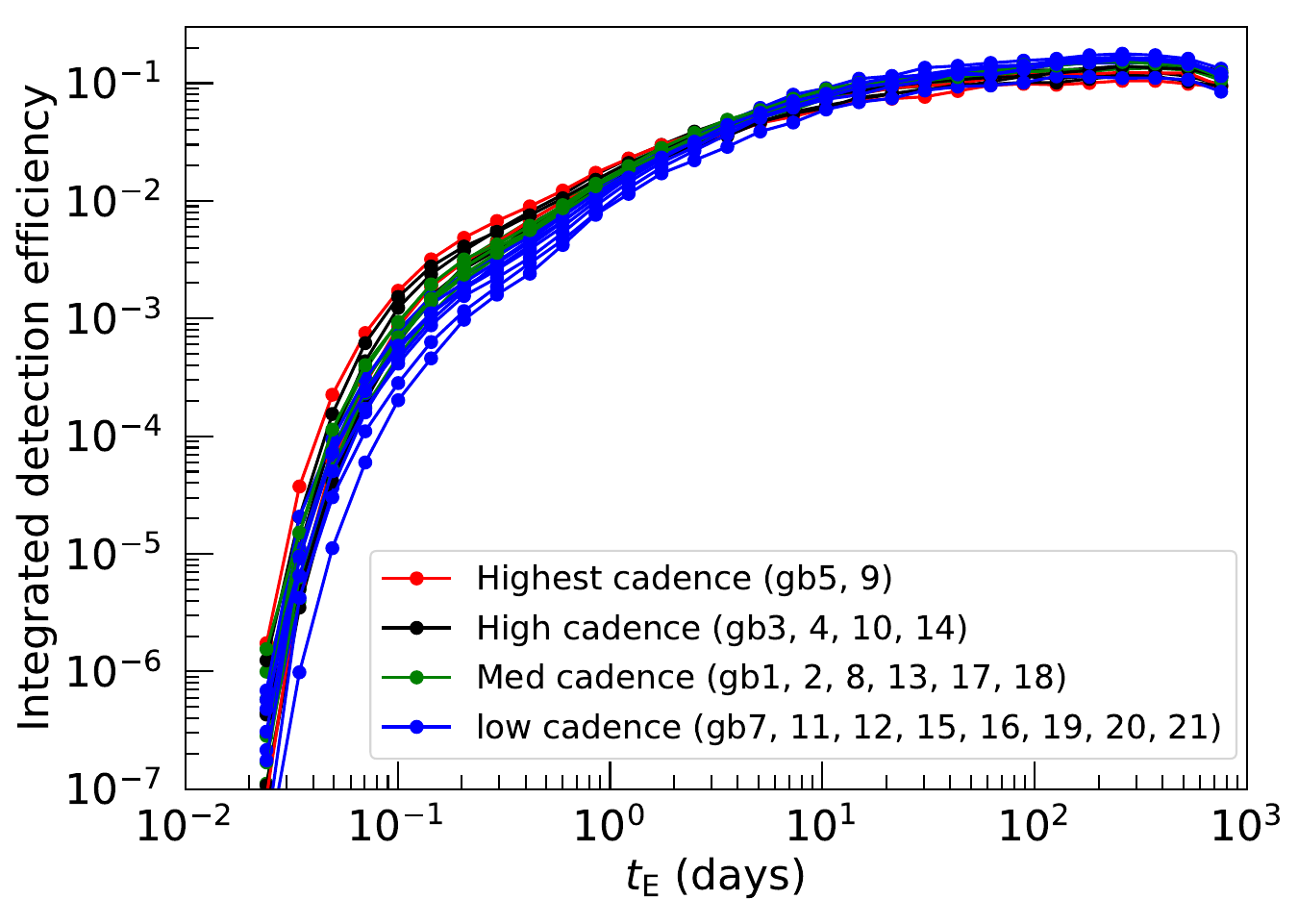}
\caption{
  \label{fig:EFF}
Integrated detection efficiencies, $\tilde{\epsilon} (t_{\rm E} ; \Gamma)$, as a function of the timescale $t_{\rm E}$ down to 
the source magnitude of $I_{\rm s} <21.4$ mag for the criteria CR2.
Red, black, green and blue lines indicate the efficiencies of fields with the highest, high, medium and low cadence, respectively.
}
\end{center}
\end{figure}

Figure \ref{fig:EFF} shows the integrated detection efficiencies $\tilde{\epsilon} (t_{\rm E} ; \Gamma)$ for the event rate calculated with the best fit MF model with the criteria CR2.
The curve for CR1 is similar.
This detection efficiency is about a factor two lower than that of \cite{Mroz17} at the low end around $t_{\rm E} = 0.1$ days 
even for the similar cadence of the survey. 
The main reason is likely that  \cite{Mroz17} did not include the finite source effect in their simulation.
\cite{Koshimoto2023} confirmed that this difference is about a factor two at $t_{\rm E} = 0.1$ days by the simulation in their Figure 7.
 
Note that detection efficiencies at short  $t_{\rm E}$ with $\rho>1$ may be improved in a future analysis.
For events with $\rho>1$, the magnification can be significant with the minimum impact parameter up to $u_{\rm 0} \lesssim \rho$.
This is likely to be more important for bright giant sources because these have higher S/N ratio even 
at low magnification (see also Appendix \ref{sec-append}) .
However such events are rejected by the criterion   $u_{\rm 0} \le 1$ in \citetalias{Koshimoto2023}.
This criterion is applied because it is useful to robustly remove the various artifacts and keep the sample as clean as possible. 
This may be improved in a future analysis with a more careful investigation.

\subsubsection{Likelihood for short timescale ($t_{\rm E} < 1$ day) events} \label{sec:L_shorttE}

We follow the importance sampling method used by \citet{hog10} to  
convert the the uninformative ``interim prior\rlap", $p_0 (\log t_{\rm E}, \log \theta_{\rm E})$ 
used for the the light curve MCMC for each event,
to a probability distribution for event $i$, ${\cal G} (\log t_{{\rm E}, i}, \log \theta_{{\rm E}, i} ; \Gamma)$, that depends 
on the event rate for each mass function model, $\Gamma$. 
However, while \citet{hog10} characterized their calculation as a modification
of the assumed prior, based on the data, this in not the case for our analysis. Instead, we are replacing $p_0$ with 
the probability distribution implied by our mass function model, using the importance sampling 
Monte Carlo integration method \citep{num_recipies}.
We use the probability distribution for each event from its MCMC analysis to
 calculate the likelihood for the short timescale events, ${\cal L}_{\rm short}$. Given the output MCMC samples of posterior distributions for individual events by \citetalias{Koshimoto2023}, the likelihood is 
given by 
\begin{align}
    {\cal L}_{\rm short} \propto \prod_{i = 1}^{N_{\rm short}} \sum_{k = 1}^{K_i} \frac{{\cal G} (\log t_{{\rm E}, ik}, \log \theta_{{\rm E}, ik} ; \Gamma)}{p_0 (\log t_{{\rm E}, ik}, \log \theta_{{\rm E}, ik})}, \label{eq:Lshort}
\end{align}
where $i$ runs over all the $N_{\rm short}$ events that have the best-fit $t_{\rm E} < 1$ day ($N_{\rm short} = 12$ for CR1 and $N_{\rm short} = 10$ for CR2), $k$ runs over all the $K_i$ samples in the MCMC sample of
the probability distribution for $i$th event, and $p_0 (\log t_{\rm E}, \log \theta_{\rm E})$ is the uninformative prior 
distribution used for these MCMC calculations.
The model's detectable event rate as a function of $(\log t_{\rm E}, \log \theta_{\rm E})$ is given by
\begin{align}
    {\cal G} (\log t_{\rm E}, \log \theta_{\rm E} ; \Gamma) = \sum_j w_j \, g_j (\log t_{\rm E}, \log \theta_{\rm E} ; \Gamma_j)
\end{align}
with
\begin{align}
    g_j (\log t_{\rm E}, \log \theta_{\rm E} ; \Gamma_j) = \epsilon_{j} (\log t_{\rm E}, \log \theta_{\rm E}) \, \Gamma_j (\log t_{\rm E}, \log \theta_{\rm E}), \label{eq:g_tEthE}
\end{align}
where we represented it as a function of $(\log t_{\rm E}, \log \theta_{\rm E})$ rather than $(t_{\rm E}, \theta_{\rm E})$ because 
the MCMC calculations of \citetalias{Koshimoto2023} provide the probability distributions 
based on the uninformative uniform prior in $(\log t_{\rm E}, \log \theta_{\rm E})$, i.e.,
$p_0 (\log t_{\rm E}, \log \theta_{\rm E}) = {\rm const.}$.

Eq. (\ref{eq:Lshort}) calculates the likelihood by summing the ratio of ${\cal G} (\log t_{\rm E}, \log \theta_{\rm E} ; \Gamma)$ to 
$p_0 (\log t_{\rm E}, \log \theta_{\rm E})$ to replace the uniform prior 
(i.e., $p_0$), used for the MCMC calculations with the new probability distribution 
(i.e., ${\cal G}$) that depends on our mass function model. 
This method, which uses all the MCMC samples, allows ${\cal L}_{\rm short}$ to account for the uncertainty of the parameters, unlike ${\cal L}_{\rm long}$ given in Eq. (\ref{eq:Llong}).

Despite the significant computational cost of Eq. (\ref{eq:Lshort}) associated with performing a summation over $K_i$ (typically $\sim 5 \times 10^5$) samples for each proposed mass function during the fitting process, we addressed this by implementing a binning strategy for the MCMC sample using grids of $(\log t_{\rm E}, \log \theta_{\rm E})$ with a size of (0.05 dex $\times$ 0.05 dex), which significantly increased the computational efficiency.

\subsection{Mass function of known population} \label{sec:MFknown}

Firstly, we perform the likelihood analysis without the short events with $t_{\rm E}< 1$ day using the Galactic model 
with the MF of known population, i.e., stellar remnants (black holes (BH), neutron stars (NS) and white dwarfs(WD)), 
main sequence stars (MS) and brown dwarfs (BD).
We use a broken power-law MF given by
\begin{equation}
 \frac{dN}{d\log M} \propto
  \left\{
   \begin{array}{ll}
     {M^{-\alpha_1}} & (M_1 < M/M_\sun < 120)\\
     {M^{-\alpha_2}} & (0.08 < M/M_\sun < M_1) \\
     {M^{-\alpha_3}} & (3 \times 10^{-4} < M/M_\sun < 0.08). \\
    \end{array}
    \right.
  \label{eq:MF}
\end{equation}

We adopt the values of parameters $\alpha_1=1.32$ and   $\alpha_2=0.13$,  $\alpha_3=-0.82$ and $ M_1=0.86$
from the E+E$_{\rm X}$ model of \citet{Koshimoto2021} by default unless specified as fitting parameters in the following three models.
The minimum mass $3 \times 10^{-4} \, M_\sun$ is taken to be smaller than 
the theoretical minimum mass of the gas cloud, $\sim$Jupiter-mass, 
that collapses to form a brown dwarf \citep{bos03}.
During our fitting procedure, a proposed initial mass function (IMF) is converted into a present-day mass function following the procedure used by \citet{Koshimoto2021} that combines their stellar age distribution and the initial-final mass relation by \citet{lam20} to evolve stars into stellar remnants.

We consider three models here: BD1, BD2, and BD3. In BD1, we fit only $\alpha_3$ as a fitting parameter, while fixing $\alpha_1$, $\alpha_2$, and $M_1$. Similarly, in BD2, we fit $\alpha_3$ and $\alpha_2$, and in BD3, we fit $\alpha_1$, $\alpha_2$, $\alpha_3$, and $M_1$. To perform the fitting, we use the Markov Chain Monte Carlo (MCMC) method \citep{metrop}, and assign uniform distributions as priors for all the parameters.

The best fit models BD1, BD2 and BD3 are almost indistinguishable from the blue dotted line in Figure \ref{fig:tE}.
One can see that the models fit the data with $t_{\rm E} > 1$ day very well.
The best fit parameters and $\chi^2$ values are listed in Table \ref{tbl:fitparamBD}.
There is no significant difference in the resultant parameters between different selection criteria or among the BD1, BD2, and BD3 models.

 All of the parameters are consistent with those of \citet{Koshimoto2021} within $1\sigma$.
This indicates that our dataset confirmed the Galactic model and MF of known objects by \citet{Koshimoto2021}. This also indicates that our dataset is consistent with the OGLE-IV $t_{\rm E}$ distribution for $t_{\rm E} > 1$ day \citep{Mroz17, Mroz19} that is fitted by \citet{Koshimoto2021}.

In the following analysis, we fit only $\alpha_3$ and 
fix all other parameters for the known populations.
Note, in  \citet{Koshimoto2021}, the Galactic model and MF are constrained to satisfy 
the microlensing $t_{\rm E}$ distribution, stellar number counts and the Galactic Bulge mass 
from other observations, simultaneously. 
In principle, the MF should not be changed alone because it is related to other parameters of the Galactic model.
However, the contribution of objects with $M/M_\sun < 0.08$ are negligible in stellar number counts and as a fraction of the Galactic Bulge mass.
Thus, we assume that a model with a different slope at lower masses with $M/M_\sun < 0.08$ is still valid.

\begin{deluxetable*}{lrrrrrrr}
\tabletypesize{\scriptsize}
\tablecaption{Best fit parameters of the mass function for known population.
 \label{tbl:fitparamBD}}
\tablewidth{0pt}
\tablehead{
\colhead{model}      &      \multicolumn{2}{c} {BD1}               &          \multicolumn{2}{c} {BD2}                 &           \multicolumn{2}{c} {BD3}                &   \citetalias{Koshimoto2021}\tablenotemark{a} \\
\colhead{} & \colhead{CR1}     &  \colhead{CR2}          &         \colhead{CR1}   &  \colhead{CR2}          & \colhead{CR1}           &  \colhead{CR2}          &
}                                                                                                                                                
\startdata
$M_1$           & ($0.86)$               &   ($0.86)$              &      ($0.86)$           &       ($0.86)$          & $0.97^{-0.04}_{-0.34}$  & $0.99^{-0.06}_{-0.37}$  &   $ 0.86^{+0.09}_{-0.10}$     \\
$\alpha_1$      & ($1.32)$               &   ($1.32)$              &      ($1.32)$           &       ($1.32)$          & $1.33^{+0.21}_{-0.17}$  & $1.34^{+0.18}_{-0.18}$  &   $ 1.32^{+0.14}_{-0.10}$     \\
$\alpha_2$      & ($0.13)$               &   ($0.13)$              & $0.20^{+0.07}_{-0.05}$  & $0.20^{+0.07}_{-0.05}$  & $0.23^{+0.04}_{-0.19}$  & $0.24^{+0.04}_{-0.21}$  &   $ 0.13^{+0.11}_{-0.12}$     \\
$\alpha_3$      &$-0.60^{+0.08}_{-0.13}$ & $-0.62^{+0.09}_{-0.14}$ & $-0.74^{+0.13}_{-0.30}$ & $-0.76^{+0.14}_{-0.30}$ & $-0.76^{+0.19}_{-0.26}$ & $-0.79^{+0.22}_{-0.25}$ &   $-0.82^{+0.24}_{-0.51}$     \\
$\chi^2$        &  35919.4               &        35722.6          &           35918.2       &       35721.5           &      35918.0            &   35721.3               &              \\
\enddata
\tablenotetext{a}{Results of fitting to various bulge data including the OGLE-IV $t_{\rm E}$ distribution of $t_{\rm E} > 1$ day \citep{Mroz17,Mroz19}. The representative values are shifted to the ones for the E+E$_{\rm X}$ model from their original ones for the G+G$_{\rm X}$ model.}
\tablecomments{Some of the upper errors of $M_1$ is negative because the best fit value is outside of the 68\% range. This is because $M_1$ is restricted to be less than 1 $M_\sun$.}
\end{deluxetable*}

\subsection{Mass function of planetary mass population} \label{sec:MFFFP}

If the candidates with $t_{\rm E}<0.5$ day are really due to microlensing, 
they can not be explained by known populations, i.e., stellar remnants, MS or BD.
To explain the tail for short values of $t_{\rm E}$, 
we defined a new model ``PL" which introduces a planetary mass population 
by the following power law in addition to known populations (Eq. \ref{eq:MF}),

\begin{equation}
  \frac{dN_4}{d\log M}  =Z  \left(\frac{M}{M_{\rm norm}}\right)^{-\alpha_4},  (M_{\rm min} < M/M_\sun<0.02).
  \label{eq:MF_PL}
\end{equation}
Here $Z$ is a normalization factor and
$M_{\rm norm}$ is a reference mass whose inclusion allows
$Z$ to have a unit of (dex)$^{-1}$. 
Although $M_{\rm norm}$ can be an arbitrary zero point, 
we found that the uncertainty in $Z$ is minimized when we adopt 
$M_{\rm norm}=8\,M_\earth$ which is recognized as a pivot point.


In the model PL, we use $\alpha_3$, $\alpha_4$ and $Z$, as fitting parameters and fix parameters
$\alpha_1=1.32$,  $\alpha_2=0.13$ and $ M_1=0.86$ \citep{Koshimoto2021}. We assign
uniform distributions as priors for $\alpha_3$, $\alpha_4$ and $\log Z$ in our MCMC run.
We found that the fitting result does not depend on $M_{\rm min}$ at all when $M_{\rm min} < 3 \times 10^{-7} \, M_\sun$, which indicates our data sensitivity is down to $\sim 3 \times 10^{-7} M_\sun$. Thus, we decided to use $M_{\rm min} = 10^{-7}$.

\begin{figure}
\begin{center}
\includegraphics[scale=0.35,keepaspectratio]{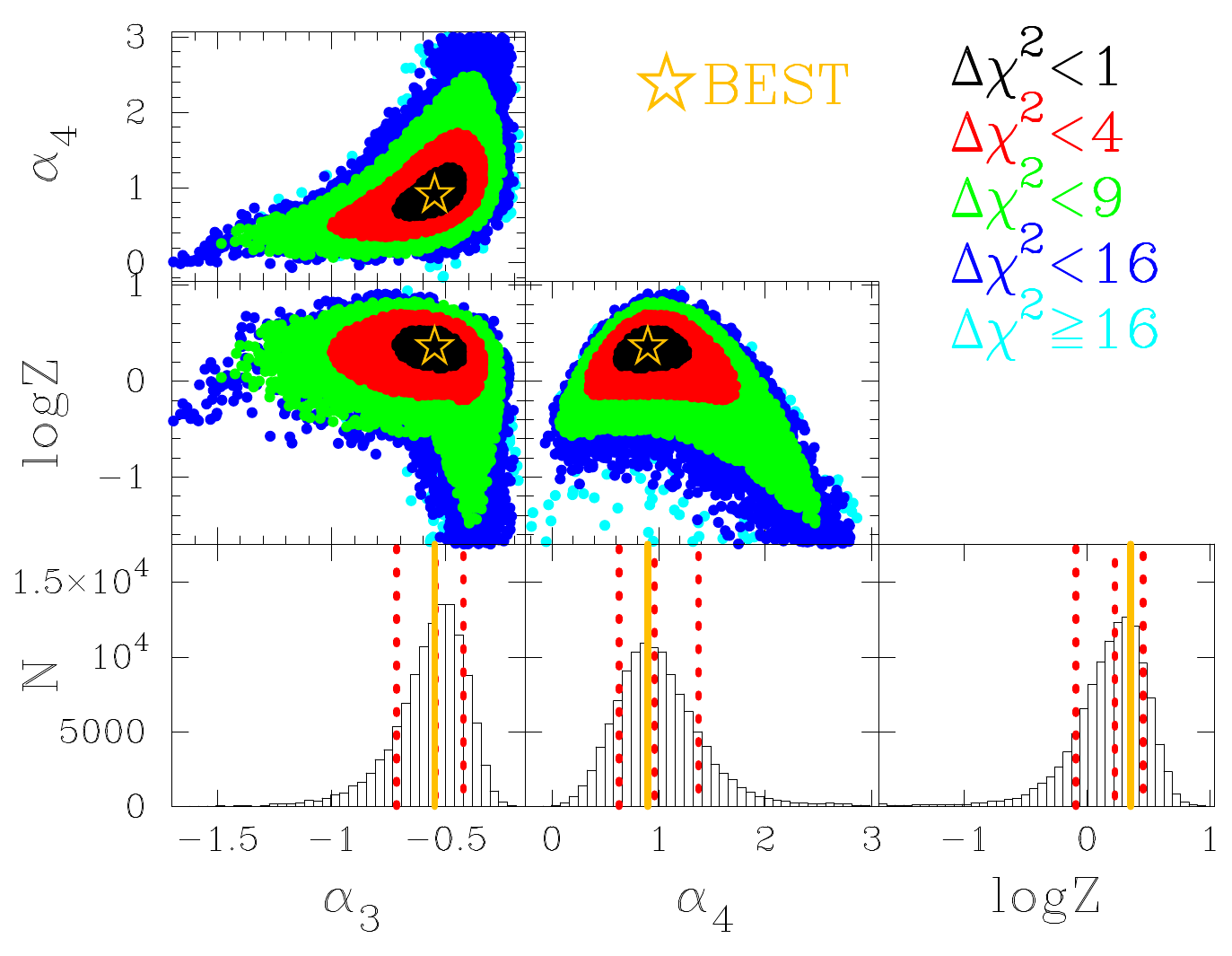}
\caption{
  \label{fig:MCMC_PL}
Posterior distributions of the parameters of the PL model for sample CR2.
The vertical red dotted lines indicate the median and $\pm 1\sigma$.
The vertical orange line indicates the best fit.
}
\end{center}
\end{figure}

The red solid line in Figure \ref{fig:tE} represents the best fit model for all populations with the CR2 sample.
This figure indicates that the model represents the observed $t_{\rm E}$ distribution well.
Note that although the observed $t_{\rm E}$ distribution shown in black in Figure \ref{fig:tE} does not include error bars along the $t_{\rm E}$ axis, the best-fit line is derived from our likelihood analysis that takes into account the $t_{\rm E}$ errors as well as the $\theta_{\rm E}$ constraints for the short events with $t_{\rm E} < 1$ day.
Figure \ref{fig:MCMC_PL} shows the posterior distributions of the parameters of PL model.
The best fit parameters and $\chi^2$ are listed in Table \ref{tbl:fitparamPL}.

The best fit power index for BD is $\alpha_3= -0.58^{+0.12}_{-0.16}$ which 
is consistent with the model without the planetary mass population.

\begin{figure}
\begin{center}
\includegraphics[scale=0.49,keepaspectratio]{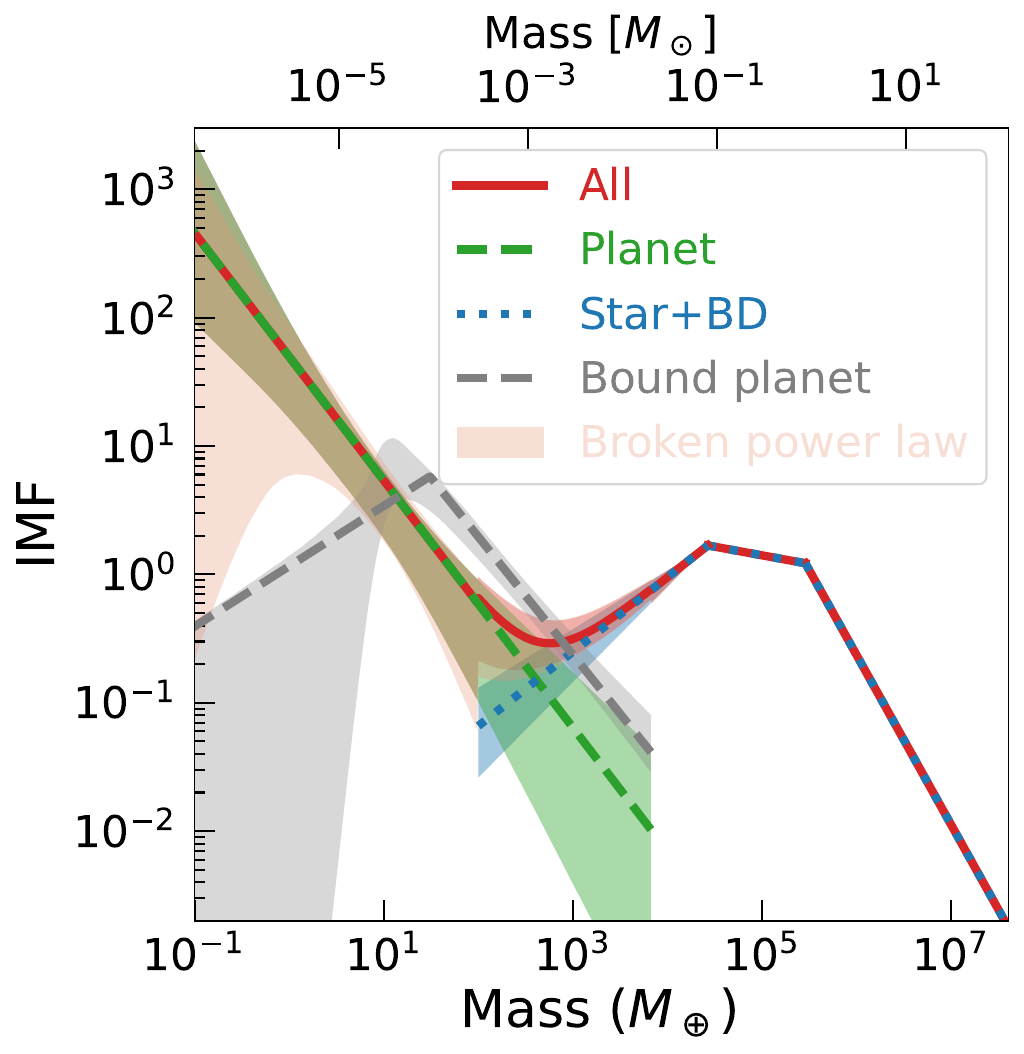}
\caption{
  \label{fig:IMF}
Initial mass function (IMF) of the best fit PL model for CR2. 
The red line indicates the best fit for all population.
The blue dotted line and green dashed line show the IMFs for the stellar and brown dwarf population and for the planetary mass population, respectively.
The shaded areas indicate $1\sigma$ error. 
The gray dashed-line and the shaded area indicate the best-fit and $1~\sigma$ range of the bound planet MF
by \citet{suzuki2016} via microlens.
The pink shaded area indicate $1\sigma$ uncertainty for the broken power law FFP model.
}
\end{center}
\end{figure}

The best fit MF of the planetary mass populations 
with the normalization $Z$ relative to stars (MS+BD+WD)  (integrated IMF over $3\times 10^{-4}<M/M_\sun < 8$)
can be expressed as  
\begin{equation}
  \frac{dN_4}{d\log M}  = \frac{2.18^{+0.52}_{-1.40}}{\rm dex\times star}   \left( \frac{M}{8\,M_\earth} \right)^{-\alpha_4}, 
  \label{eq:MF_PL2}
\end{equation}
where  $\alpha_4= 0.96^{+0.47}_{-0.27}$.
Figure \ref{fig:IMF} shows the IMF of the best fit PL model.
This $\alpha_4$ is consistent with the corresponding power law index 
of $0.9\lesssim p\lesssim1.2$ suggested by \cite{Gould2022}.

This can be translated to the normalization per stellar mass
of stars, $Z^{M_\sun}$, as,

\begin{equation}
  \frac{dN_4}{d\log M} =\frac{5.48^{+1.18}_{-3.50}}{ {\rm dex} \times M_\sun}  \left( \frac{M}{8\,M_\earth} \right)^{-\alpha_4}.
  \label{eq:ZM_MS_BD}
\end{equation}
This implies that the number of FFPs per stars is $f= 21^{+23}_{-13}$\,star$^{-1}$
over the mass range $10^{-6}<M/M_\sun < 0.02$ ($0.33<M/M_\earth < 6660$).
Note that this value is vary depending on the minimum mass.
The total mass of FFPs per star is $m =  80^{+73}_{-47} M_\earth (0.25_{-0.15}^{+0.23} M_{\rm J})$\,star$^{-1}$.
This is less dependent from the minimum mass.
The total mass of FFPs per $M_\sun$ is  
$m^{M_\sun}=202^{+166}_{-114}$ $M_\earth (0.64_{-0.11}^{+0.19} M_{\rm J}) M_\sun^{-1}$. 
This is more robust values less dependent on uncertainty in the abundances 
of the low mass objects for both FFP and BD.

The normalization, number and total mass of FFP relative to MS+BD ($3\times 10^{-4}<M/M_\sun < 1.1$) are 
also shown in Table \ref{tbl:fitparamPL}.
These normalizations can be translated to 
$Z_{\rm MS+BD}=0.53^{+0.19}_{-0.40}$ dex$^{-1}$star$^{-1}$ and 
$Z_{\rm MS+BD}^{M_\sun}=2.44^{+0.71}_{-1.82}$ dex$^{-1} M_\sun ^{-1}$
with $M_{\rm norm}=38M_\earth$.
These are almost same as
$Z_{\rm MS+BD}=0.39\pm0.18$ dex$^{-1}$star$^{-1}$ and
$Z_{\rm MS+BD}=1.96\pm0.98$ dex$^{-1}M_{\sun}^{-1}$
with $M_{\rm norm}=38M_\earth$
by \cite{Gould2022}.


Note that the lenses for these short events could be either FFP or planets 
with very wide separations of more than about ten astronomical units (AU) from their host stars, 
for which we cannot detect the host star in the light curves.

%

\begin{deluxetable}{lccccc}
\tabletypesize{\scriptsize}
\tablecaption{Best fit parameters of the mass function for the planetary mass population.
 \label{tbl:fitparamPL}}
\tablewidth{0pt}
\tablehead{
                           &   \colhead{CR1}          &          \multicolumn{2}{c}{CR2}                     &     \citetalias{Gould2022}         \\
\colhead{($M_{\rm norm}$)} & \colhead{($8~M_\earth$)} & \colhead{($8~M_\earth$)} & \colhead{($38~M_\earth$)} & \colhead{($38~M_\earth$)}
}
 \startdata
 $M_1$                           & ($0.86)$                    &       ($0.86)$          &                         &                   \\
 $\alpha_1$                      & ($1.32)$                    &       ($1.32)$          &                         &                   \\
 $\alpha_2$                      & ($0.13)$                    &       ($0.13)$          &                         &                   \\
 $\alpha_3$                      & $-0.55^{+0.13}_{-0.17}$     & $-0.58^{+0.12}_{-0.16}$ &                         &                   \\
 $\alpha_4$                      &  $0.90^{+0.48}_{-0.27}$     &  $0.96^{+0.47}_{-0.27}$ &                         &  fixed at $0.9$ or $1.2$       \\
 $Z$                       & $2.08^{+0.54}_{-1.33}$ & $2.18^{+0.52}_{-1.40}$  & $0.49^{+0.17}_{-0.37}$  &                   \\
 $Z_{\rm MS+BD}$                 &  $2.27^{+0.60}_{-1.46}$     &  $2.38^{+0.58}_{-1.53}$ & $0.53^{+0.19}_{-0.40}$  &  $0.39 \pm 0.20 \pm ?$  \\
 $Z^{M_\sun}$              & $5.33^{+1.26}_{-3.40}$  & $5.48^{+1.18}_{-3.50}$ & $1.22^{+0.35}_{-0.91}$  &                   \\
 $Z^{M_\sun}_{\rm MS+BD}$        & $10.63^{+2.52}_{-6.78}$     & $10.95^{+2.36}_{-6.97}$ & $2.44^{+0.71}_{-1.82}$  &  $1.96 \pm 0.98 \pm ?$  \\
 $f$\tablenotemark{a}             &  $17^{+20}_{-11}$       & $21^{+23}_{-13}$       &                   &                   \\
 $f_{\rm MS+BD}$\tablenotemark{a} & $19^{+22}_{-12}$              &  $23^{+25}_{-15}$            &                   &                   \\
 $f^{M_\sun}$\tablenotemark{a} &  $45^{+54}_{-30}$    & $53^{+59}_{-34}$       &                   &                   \\
 $f^{M_\sun}_{\rm MS+BD}$\tablenotemark{a} & $89^{+107}_{-59}$    &  $106^{+117}_{-68}$          &                   &                   \\
 $m$\tablenotemark{b}             & $89^{+96}_{-56} M_\earth$  & $80^{+73}_{-47} M_\earth$  &                   &                   \\
 $m_{\rm MS+BD}$\tablenotemark{b} &  $98^{+107}_{-61} M_\earth$   &  $88^{+81}_{-51} M_\earth$   &                   &                   \\
 $m^{M_\sun}$\tablenotemark{b} & $229^{+219}_{-140} M_\earth$ & $202^{+166}_{-114} M_\earth$      &           &                   \\
 $m^{M_\sun}_{\rm MS+BD}$\tablenotemark{b} & $457^{+439}_{-279} M_\earth$ & $404^{+333}_{-228} M_\earth$ &           &                   \\
 $\chi^2$                       &   36273.0                       &  36024.1                &                        &                   \\
 \enddata
\tablenotetext{a}{Number of planetary mass objects per BD+MS+WD ($f$), per MS+BD ($f_{\rm MS+BD}$), per solar mass of BD+MS+WD ($f^{M_\sun}$) or per solar mass of MS+BD ($f^{M_\sun}_{\rm MS+BD}$) when MF down to $10^{-6} M_\sun$ are integrated. These are vary depending on the minimum mass.}
\tablenotetext{b}{Total mass of planetary mass objects per BD+MS+WD ($m$), per MS+BD ($m_{\rm MS+BD}$), per solar mass of BD+MS+WD ($m^{M_\sun}$) or per solar mass of MS+BD ($m^{M_\sun}_{\rm MS+BD}$) when MF down to $10^{-6} M_\sun$ are integrated.}
\tablecomments{We adopt the model for CR2 as the final result.}
\end{deluxetable}

\subsection{Broken power law MF for the planetary mass population} \label{sec:MFFFP_broken}
In order to demonstrate the FFP mass function uncertainty at low masses, we
have also modeled the planetary mass population with a broken power law MF given by
\begin{equation}
 \frac{dN_4}{d\log M} \propto
  \left\{
   \begin{array}{ll}
     Z  \left(\frac{M}{M_{\rm norm}}\right)^{-\alpha_4},  & (M_{\rm br} < M/M_\sun<0.02)\\
     M^{-\alpha_5}, & (M_{\rm min} < M/M_\sun < M_{\rm br}). \\
    \end{array}
    \right.
  \label{eq:MFFFP_broken}
\end{equation}
Here, $M_{\rm br}$ is a break mass and $\alpha_5$ is a power bellow $M_{\rm br}$.
$M_{\rm min} = 10^{-7}M_\sun$ is same as the previous section.

In Figure \ref{fig:IMF}, we show the  1 $\sigma$ range of the broken power law PL model
along with the best fit single power law MF given in the previous section for comparison.
The median and 1 $\sigma$  range of the parameters and $\chi^2$ are listed in Table \ref{tbl:fitparamPL_broken}.
The resultant broken power law MF is consistent with the single power law model
while the uncertainty is larger.
Although the MF is relatively well constrained down to an Earth mass,
the uncertainty is much larger bellow an Earth mass. 
This is as expected because of our low sensitivity bellow to planets of less than an Earth mass.

This model implies that the number of FFPs per stars is  $f=17^{+39}_{-12}$\,star$^{-1}$
over the mass range $10^{-6}<M/M_\sun < 0.02$ ($0.33<M/M_\earth < 6660$).
The total mass of FFPs per star is $m =  69^{+107}_{-36}$ $M_\earth (0.22_{-0.11}^{+0.33} M_{\rm J})$\,star$^{-1}$.
The total mass of FFPs per $M_\sun$ is  
$m^{M_\sun}=175^{+246}_{-89}$ $M_\earth (0.55_{-0.28}^{+0.77} M_{\rm J}) M_\sun^{-1}$. 
These numbers are also consistent with those for the single power law model but has larger uncertainties.
This result is useful to see the conservative uncertainty of MF.
In the following discussion, although we use only the results for the single power law model,
the discussion is qualitatively same for the broken power law.


\begin{deluxetable}{lccc}
\tabletypesize{\scriptsize}
\tablecaption{Median and uncertainty of  parameters of the broken power law mass function for the planetary mass population.
 \label{tbl:fitparamPL_broken}}
\tablewidth{0pt}
\tablehead{
                           &   \colhead{CR1}          &          \multicolumn{1}{c}{CR2}                         \\
\colhead{($M_{\rm norm}$)} & \colhead{($8~M_\earth$)} & \colhead{($8~M_\earth$)} 
}
 \startdata
                               $\alpha_3$ &        $-0.54^{+0.12}_{-0.17}$ &        $-0.58^{+0.12}_{-0.19}$ \\
                               $\alpha_4$ &        $ 1.07^{+0.93}_{-0.49}$ &        $ 1.14^{+0.97}_{-0.54}$ \\
                               $\alpha_5$ &        $ 0.13^{+1.33}_{-3.07}$ &        $ 0.13^{+1.32}_{-3.10}$ \\
                        $\log M_{\rm br}$ &        $-5.35^{+1.35}_{-1.02}$ &        $-5.27^{+1.28}_{-1.05}$ \\
                                      $Z$ &        $ 1.79^{+2.91}_{-1.08}$ &        $ 1.85^{+3.14}_{-1.17}$ \\
                          $Z_{\rm MS+BD}$ &        $ 1.96^{+3.19}_{-1.18}$ &        $ 2.03^{+3.43}_{-1.28}$ \\
                             $Z^{M_\sun}$ &        $ 4.57^{+7.54}_{-2.77}$ &        $ 4.62^{+7.92}_{-2.91}$ \\
                 $Z^{M_\sun}_{\rm MS+BD}$ &       $ 9.12^{+15.05}_{-5.52}$ &       $ 9.22^{+15.79}_{-5.82}$ \\
                     $f$\tablenotemark{a} &               $15^{+36}_{-11}$ &               $17^{+39}_{-12}$ \\
         $f_{\rm MS+BD}$\tablenotemark{a} &               $17^{+40}_{-12}$ &               $18^{+42}_{-13}$ \\
            $f^{M_\sun}$\tablenotemark{a} &               $39^{+96}_{-28}$ &               $42^{+98}_{-30}$ \\
$f^{M_\sun}_{\rm MS+BD}$\tablenotemark{a} &              $79^{+191}_{-57}$ &              $85^{+196}_{-61}$ \\
                                      $m$ &              $73^{+119}_{-40}$ &              $69^{+107}_{-36}$ \\
         $m_{\rm MS+BD}$\tablenotemark{b} &              $80^{+131}_{-44}$ &              $75^{+118}_{-39}$ \\
            $m^{M_\sun}$\tablenotemark{b} &            $192^{+275}_{-103}$ &             $175^{+246}_{-89}$ \\
$m^{M_\sun}_{\rm MS+BD}$\tablenotemark{b} &            $384^{+551}_{-206}$ &            $349^{+493}_{-178}$ \\
                $\chi^2$                       &   36271.6                       &  36022.9                                          \\
  \enddata
\tablenotetext{a}{Same as Table \ref{tbl:fitparamPL} }
\tablenotetext{b}{Same as Table \ref{tbl:fitparamPL}}
\tablecomments{The median and 1$\sigma$ ranges are shown for understanding the uncertainty.}
\end{deluxetable}


%

\subsection{Comparison to  \cite{sumi2011}} \label{sec:Comparison2Sumi11}
As discussed in \citetalias{Koshimoto2023}, the data reduction for the MOA-II 9-year 
analysis was done using an improved data reduction method, with the primary improvement being
the introduction of a photometry detrending method introduced by \cite{bennett2012} and used by
\cite{sumi2016}. This method is able to largely remove systematic errors due to color-dependent atmospheric refraction
that can shift the position of neighbor stars of different colors towards or away from the target star
as star rises and sets, as discussed in Section~\ref{sec:Data}. 
This systematic error due to atmospheric refraction could cause light curve
variations on a daily timescale, and these were the likely cause of the feature at $t_{\rm E} \sim 1\,$day
in the MOA-II 2-year analysis that was attributed by \cite{sumi2011} (S11 hereafter) to a large number
of FFPs with masses similar to Jupiter's mass. This was based on 10 events 
with $0.5 < t_{\rm E}/{\rm day} < 2$.

Our new analysis of the 9-year data set has found fewer 2006 and 2007 events with 
$0.5 < t_{\rm E}/{\rm day} < 2$ than the 10 events found by S11. We find 5 such events
for selection criteria CR2, with one additional event
passing selection criteria CR1. Two of these $0.5 < t_{\rm E}/{\rm day} < 2$ events had their best fit source magnitudes
decrease to fainter than our limit of $I_s \leq 21.4$ and their best fit $ t_{\rm E}$ values increase to $>2\,$days. Two 
other events had their $t_{\rm E}$ error bars increase to above our threshold. Both of these effects are likely
to be due to the new photometry detrending correction. Another of the 10 S11 events with $0.5 < t_{\rm E}/{\rm day} < 2$ saw
its best fit $u_0$ value increase from 0.91 to 1.01, so as to fail our $u_0 \leq 1.0$ cut, but another
$0.5 < t_{\rm E}/{\rm day} < 2$ event from the 2006-2007 time period, MOA-9y-3036, was added to the sample.
The full 9-year data set contains 
15 events with $0.5 < t_{\rm E}/{\rm day} < 2$, which is $3\times$ less than rate predicted by the 2-year S11 analysis. 
This is largely explained by our detrending routine which increased the $t_{\rm E}$ values for some short
events and reduced the estimated $t_{\rm E}$ measurement precision for other short events.

An additional, shorter event with $ t_{\rm E} < 0.5\,$days, MOA-9y-6057, from 2006, was also found in the
9-year analysis, but this event was not found in the S11 analysis. The full 9-year sample has 6 events with 
$ t_{\rm E} < 0.5\,$days, including 2 with finite source effects that were not considered in the S11 analysis.
The lack of such events in the S11 analysis is largely due to Poisson statistics, since the two events that could
have failed the S11 event selection due to finite source effects did not occur in the two years of the S11 sample.

The number of events predicted to be found in the $0.5 < t_{\rm E}/{\rm day} < 2$ range has also changed
for reasons relating to our light curve analysis, but changes to our Galactic model may have had a more 
significant effect. The systematic errors that were largely corrected by our detrending method had
the most significant effect on events with $t_{\rm E} \sim 1\,$day. This systematic error inflated the 
number of events in the $0.5 < t_{\rm E}/{\rm day} < 2$ range in S11, but the  also reduced the number of 
events in the $2 < t_{\rm E}/{\rm day} < 4$ range. This resulted 
in an underestimation of the number of brown dwarfs by pushing the brown dwarf power law to $\alpha_3= -0.5 $,
and this inflated the number of Jupiter-mass FFPs needed to explain the events in the $0.5 < t_{\rm E}/{\rm day} < 2$ range.
The model found in the \cite{Mroz17} analysis, which was based on the higher quality OGLE light curves 
predicted more brown dwarfs than S11 with a slope of $\alpha_3=-0.2$, which greatly reduced
the FFP contribution needed to explain events in the $0.5 < t_{\rm E}/{\rm day} < 2$ range.

Much of the change in the interpretation of events in the $0.5 < t_{\rm E}/{\rm day} < 2$ in our 9-year analysis
came from changes in the Galactic model used. The 9-year analysis uses the \cite{Koshimoto2021} Galactic
model, which has been specifically designed to match the Galactic properties, such as proper motion distributions
that are the most important for the interpretation of microlensing events. This new Galactic model increases 
the width of the $t_{\rm E}$ distribution for lenses of a fixed mass by $\sim 24$\%, and this led to an increase
in the number of main sequence stars and brown dwarfs contributing to the number of $0.5 < t_{\rm E}/{\rm day} < 2$
events. Also, the S11 model cut off the brown dwarf mass distribution at $0.01M_\sun$, whereas we have
extended this cutoff down to $3\times 10^{-4}M_\sun$ in this 9-year analysis. These changes increased 
the number of brown dwarfs, although the best fit slope $\alpha_3 = -0.58$ of the brown dwarf mass function
is similar to the S11 value.

Our best fit model for the 9-year sample now includes the following lens contributions to the events in the
$0.5 < t_{\rm E}/{\rm day} < 2$ range: 2.0 main sequence stars, 12.9 brown dwarfs (including 4.9 with $M < 0.01 M_\sun$),
and 3.6 FFP, for a total of 18.5 events. The favored model of S11, extended to a 9-year survey, would predict
0.9 main sequence stars, 4.4 brown dwarfs, and 39.7 FFP, for a total of 45.0 events. So, the new model predicts
59\% fewer events than the S11 model in the $0.5 < t_{\rm E}/{\rm day} < 2$ range, and only 19.5\% of these
events are due to FFP, compared to 88.2\% in the S11 model.

\section{Discussion and conclusions}
\label{sec:discussionAndSummary}

We derived the MF of lens objects from the 9-year MOA-II survey towards the Galactic Bulge.
The 3,535 high quality single lens light curves used in our statistical analysis include 
10 very short ($t_{\rm E}<1$ day) events, and  13 events with strong finite source
effects that allow the determination of the angular Einstein radius, $\theta_{\rm E}$.

The cumulative $\theta_{\rm E}$ histogram for these 13 events reveals an
``Einstein gap" at $5<\theta_{\rm E}/{\rm \mu as}<70$ which is roughly consistent with 
the gap at  $10<\theta_{\rm E}/\mu {\rm as} <30$ found by the KMTNet group
 \citep{Ryu2021,Gould2022}. 
This gap indicates that there is a distinct planetary mass population
separated from the known populations of brown dwarfs, stars and stellar remnants.

We constructed the $t_{\rm E}$ distribution of all selected samples including both PSPL and FSPL.
We calculated the integrated detection efficiency $\tilde{\epsilon} (t_{\rm E} ; \Gamma)$ of the survey 
by integrating the two dimensional detection efficiency, $\epsilon (t_{\rm E}, \theta_{\rm E})$,
measured from image level simulations that included the FS effect, and convolving this with
the event rate $\Gamma (t_{\rm E}, \theta_{\rm E})$ given by a Galactic model and MF.
We found that the $t_{\rm E}$ distribution has an excess at short $t_{\rm E}$ values
which can not be explained by known populations.

We then adopted the single power law MF for the planetary mass population.
We found that these short events can be well modeled by
 $dN_4/d\log M = (2.18^{+0.52}_{-1.40})\times   (M/8\,M_\earth)^{-\alpha_4}$ dex$^{-1}$star$^{-1}$ with 
 $\alpha_4 = 0.96^{+0.47}_{-0.27}$ at $10^{-7}<M/M_\odot < 0.02$ (or $ 0.033 < M/M_\earth < 6660$).

This can also be expressed by the MF per stellar mass as,
$dN_4/d\log M =5.48^{+1.18}_{-3.50}\times   (M/8\,M_\earth)^{-\alpha_4}$ dex$^{-1} M_\sun ^{-1}$.
We showed the number of FFP or distant planets is $f= 21^{+23}_{-13}$ per stars.
Note we found $f=17^{+39}_{-12}$ FFP per star for the broken power law model, 
which is consistent with our result for the single power law model, with a larger larger uncertainty.
In the following discussion, we only use the results for the single power law model,
the conclusions are qualitatively the same same for the broken power law model..

It is well known that planet-planet scattering during the planet formation process 
is likely to produce a population of
unbound or wide orbit planetary mass objects \citep{rasio96,weiden96,lin97}.
The probability of planet scattering likely increases with declining mass 
because planets usually require more massive planets to scatter.
 So, we expect the power law index of MF of bound planets $\alpha_{\rm b}$ is smaller 
 than that of $\alpha_4$ for unbound or large orbit planets, i.e., $\alpha_4 > \alpha_{\rm b}$.
 
One can compare our FFP result to the MF of known bound planets. At present, 
microlensing surveys have only measured the mass ratio function, rather than the 
mass function, of the bound planets. Currently, the most sensitive study of the bound planet
mass ratio function
\cite{suzuki2016} found that the mass ratio function can be well explained by the broken power law with
$\alpha_{\rm b}=0.93\pm0.13$ for $q> q_{\rm br} = 1.7\times 10^{-4}$,
$\alpha_{\rm b}=-0.6_{-0.5}^{+0.4}$ for $q< q_{\rm br} = 1.7\times 10^{-4}$. While the
\cite{suzuki2016} data could establish the existence of the power-law break
with reasonably high confidence (a Bayes factor of 21), there was a large, correlated
uncertainty in the mass ratio of the break and slope of the mass ratio function below the
break. So, we chose to fix the mass ratio of
break at $q_{\rm br} = 1.7\times 10^{-4}$ in order to estimate the
power law below the break. 

More recently, several papers have attempted to
improve upon this estimate by including a heterogeneous set of lower mass ratio planets
found by a number of groups without a calculation of the detection efficiency. 
These efforts included attempts to estimate the effect of a ``publication bias" 
that might cause planets deemed to be of greater interest to be published much more quickly, 
leading to biased, inhomogeneous sample of planets. This ``publication bias"  is caused by
the decision to publish some planet discoveries at a higher priority than others.
With such an analysis
\cite{Udalski2018} reported 
$\alpha_{\rm b}=-1.05_{-0.78}^{+0.68}$ with their sample and 
$\alpha_{\rm b}=-0.73_{-0.34}^{+0.42}$ when combined with the \cite{suzuki2016} result
for $q< 1\times 10^{-4} < q_{\rm br}$. A similar analysis by \citet{Jung2019}, attempted a
new measurement of the location of the break and found
$\alpha_{\rm b}=-4.5$ for $q< q_{\rm br} = 0.55\times 10^{-4}$ which is consistent
with the \cite{suzuki2016}  result when $q_{\rm br}$ is not fixed. However, a more recent
paper \citep{zang22} by many of the same authors, reported a number of planetary
microlensing events that were missed by the analyses described in \cite{Udalski2018} and 
\citet{Jung2019}. This casts some doubt on the validity of some of the assumptions
in these papers. This later paper also suggests that planets with mass ratios of $q< q_{\rm br} = 1.7\times 10^{-4}$
may be more common than previously thought, although a more definitive claim awaits a
detection efficiency calculation. Also, the \cite{suzuki2016} analysis does not imply that there is a peak
in the mass ratio. Instead it concludes that the slope does not rise as steeply toward low mass ratios
as is does for $q > 1.7\times 10^{-4}$.




The broken power-law model of  \cite{suzuki2016} is consistent with the hypothesis that these unbound or wide 
orbit planetary mass objects are the result of scattering from bound planetary systems.
It is the lower mass planets that are preferentially removed by planet-planet scattering interactions, 
so the initial planetary mass function may have been closer to a single power-law with $\alpha_{\rm b} \sim 0.9$, but 
planet-planet scattering has likely depleted the numbers of low-mass planets at separations beyond
the snow line where microlensing is most sensitive. Thus, planet-planet scattering may be responsible for the
mass ratio function ``break" observed in the \cite{suzuki2016} sample

This idea that planet-planet scattering is responsible for a FFP mass function slope
that is steeper than the slope of the mass ratio function for low-mass bound planets is also
consistent with the single power-law models that were found in smaller data sets \citep{sumi2010}.
The best fit single power-law model for the \cite{suzuki2016} sample gives 
$dN_{\rm bound}/d\log q = 0.068^{+ 0.016}_{-0.014} {\rm dex^{-2}star^{-1}}\times (q/0.001)^{-\alpha_{\rm b}}$ with
$\alpha_{\rm b}=0.58\pm 0.08$ for $3\times 10^{-6} < q< 3\times 10^{-2}$, but the broken power-law
is a significantly better fit to the  \cite{suzuki2016} data.
Note, this single power law model with $\alpha_{\rm b}=0.58$ satisfies $\alpha_4 > \alpha_{\rm b}$, for
our value of $\alpha_4 = 0.96^{+0.47}_{-0.27}$, implying that unbound (or very wide orbit) planets
increase more rapidly than bound planets at low masses.
Thus our main conclusion discussed bellow with the broken power law model, 
which the lower mass planets are increasingly scattered, is not specific to the \citet{suzuki2016} broken
power-law model.


As a comparison, we transformed the bound planet's mass ``ratio" function
of \cite{suzuki2016} to a mass function by using 
the estimated average mass of their hosts of $\sim 0.56M_\sun$ 
as shown\footnote{The 1$\sigma$ range indicated by the gray shaded area in Figure \ref{fig:IMF} does not match the one provided in \citet{suzuki2016}. This was due to an error in the \citet{suzuki2016} figure, but there is no error in the other results in that paper.}
in Figure \ref{fig:IMF}.
We estimate the abundance of the wide-orbit bound planets to be
$f_{\rm wide}=1.1_{-0.3}^{+0.6}$ planets star$^{-1}$
 in the mass range $10^{-6} < M/M_\sun < 0.02$ ($0.33<M/M_\earth < 6660$) and 
 separation range $0.3 < s < 5$, 
 which corresponds to a semi-major axis of roughly $0.7<a/{\rm au}<12$.
 This indicates that the abundance of FFP, $f=21^{+23}_{-13}$
planets star$^{-1}$, is  $19_{-13}^{+23}$
times more than wide-orbit bound planets in this mass range.

This is because the number of wide-orbit bound planets decreases at lower masses than the break
at  $M_{\rm break}\approx 1.0\times10^{-4} M_\sun$, 
while the number of high-mass bound planets is larger than that for FFP.
Again, this is consistent with the hypothesis that the low-mass planets 
are more likely to be scattered.
Note that there is still large uncertainty in the MF 
at low masses for both bound and unbound planets. It is very important to constrain the 
these MFs at low masses.

We can also compare our number for the FFP abundance with
the abundance of the bound planets with short period orbits of
$P=0.5-256$ days and planetary radii of $R_{\rm p}=0.5-4 R_\earth$
found by Kepler. \cite{Hsu2019} find
$f_{\rm FGK}= 3.5_{-0.6}^{+0.7}$ for FGK dwarfs  and \cite{Hsu2020} find
$f_{\rm M}  = 4.2_{-0.6}^{+0.6}$ for M dwarfs.
Because the typical spectral types of their samples are 
G2 ($M=1M_\sun$) and M2.5 ($M=0.4M_\sun$),
their typical semi-major axis are $0.012\lesssim a/{\rm au} \lesssim 0.79$ and
$0.009 \lesssim a/{\rm au} \lesssim 0.58$, respectively.
The fraction of FGK and M dwarfs relative to all population except BH and NS 
are  $0.157 : 0.465$ in our best fit MF.
By weighting with these stellar type fractions, the abundance of the known close-orbit bound planets is about  
$f_{\rm close}= 2.5_{-0.3}^{+0.3}$ per star. (This ignores the relatively small number of gas
giant planets in short period orbits \citep{bryant23}).

The total abundance of the wide-orbit and known close-orbit bound planets is about  
$f_{\rm bound}= 3.6_{-0.4}^{+0.7}$ per star. 
This indicates that the abundance of FFP, $f=21^{+23}_{-13}$
planets star$^{-1}$, is 
$5.8_{-3.8}^{+6.4}$ 
times more than known bound planets in this mass range.

We found the total mass of FFPs or distant planets per star 
is $m =  80^{+73}_{-47} M_\earth (0.25_{-0.15}^{+0.23} M_{\rm J})$ star$^{-1}$ in this
$10^{-6} < M/M_\sun < 0.02$ ($0.33<M/M_\earth < 6660$) mass range.
This is comparable to the value of
$91_{-22}^{+33}$ $M_\earth$ star$^{-1}$
for wide-orbit bound planets with separations of $0.3 < s < 5$ in the same mass range.
It is not straight forward to estimate the total mass of inner planet found by Kepler because only a small,
and somewhat biased, sample of Kepler planets have mass measurements.
The total masses of FFP and bound planets are less dependent on the uncertainty of the 
number of low mass planets than the total numbers of FFP and bound planets are.

These comparisons indicate that 
$19_{-13}^{+23}$
times more
planets than the ones currently in wide orbits 
have been ejected to unbound or very wide orbits.
These comparisons also suggest that the total mass of 
scattered planets is of the same order as those remaining bound
in wide orbits (beyond the snow line) in their planetary systems.
The low mass bound planets in wide orbits are much less abundant than those orbiting
closer to their host stars.
This may be explained by that planets in wide orbits are more easily ejected than those in 
close orbit.

The power-law index of the IMF of planets formed in wide orbits in 
protoplanetary disks is likely to be $\alpha_4\sim 0.9$ 
with an abundance of 
$22_{-13}^{+23}$ planets star$^{-1}$
or 
$171_{-52}^{+80} M_\earth(0.54_{-0.16}^{+0.25} M_{\rm J})$ star$^{-1}$.

Various formation mechanism of FFPs from bound planetary systems have been proposed.
Planets can be ejected from their hosts by a dynamical interactions with other (mostly giant) 
planets \citep{rasio96,weiden96,lin97}, by stellar flybys \citep{Malmberg2011},
or by the post-main-sequence evolution of their hosts \citep{Adams2013}.
\cite{Coleman2023} simulated the circumbinary planetary systems for the Kepler-16 and Kepler-34, 
and found that such systems may eject 6.3 and 9.3 planets on average, respectively, and most of these have masses
smaller than Neptune.
However, there are very few or almost no studies on the prediction for the number of the ejection of the 
Earth-Neptune mass planet population, because the abundance of such planets in less tightly bound
wide orbits is not well known. 
The results of our study may shed light on this area.

Another, rather speculative, possibility is that most of the low-mass objects
found by microlensing are primordial black holes (PBH) \citep{Niikura2019a, Niikura2019b}.
\cite{Hashino2022} predicted PBH generated at a first order electroweak phase transition
have masses of about $10^{-5} M_\sun$. They found that depending on parameters 
of the phase transition a sufficient number of PBH can be 
produced to be observed by current and future microlensing surveys.
The mass of such PBH is a function of the time of their generation, i.e., 
the electroweak phase transition, and is expected to be a  delta-function distribution.
To differentiate PBH from FFP, we need to measure the shape of 
the MF accurately.
This can be done by the current (MOA, OGLE, KMTNet) surveys, 
the near future (PRIME) ground telescope and the Roman Space telescope.

For the first time, we have determined the detection efficiency as a function of both the Einstein
radius crossing time and the angular Einstein radius, because finite source effects
have a large influence on the detectability of microlensing events due to low-mass planets.
This method is necessary for reliable results for low-mass FFPs, and it should be very useful 
for the analysis of these future surveys which will detect many short events.

A precise measurement of the free floating planet mass function will require a microlensing survey
that can obtain precise photometry of main sequence stars with relatively low magnification, because
the small angular Einstein radii, $\theta_{\rm E}$, of low-mass planetary lenses prevent high magnification.
The exoplanet microlensing survey of the Roman Space Telescope is such a survey, and it should
provide the definitive measurement of the free floating planet mass function.
\cite{Johnson2020} predicted the $\sim$250 FFPs with masses down to that of Mars 
(including $\sim$25 with masses of $0.1 \le M/M_\earth \le 1$, and $\sim 48$ with $0.316 \le M/M_\earth \le 3.16$)
assuming the fiducial mass function of cold, bound planets adapted from \cite{Cassan2012}.
Our FFP mass function results imply a large increase in the number of FFP events that should be
detected by Roman. We predict 
$988^{+1848}_{-566}$   FFPs with masses down to that of Mars (including
$575^{+1733}_{ -424}$  with $0.1 \le M/M_\earth \le 1$, and
$391^{+344}_{-259}$ with $0.316 \le M/M_\earth \le 3.16$), for our single power law model. The
broken power law model predicts 
$699^{+1424}_{ -418}$ FFPs down to that of Mars (including 
$303^{+1268}_{-271}$ with $0.1 \le M/M_\earth \le 1$, and
$261^{+436}_{ -213}$ with $0.316 \le M/M_\earth \le 3.16$).

The Earth 2.0 (ET) mission is a proposed space telescope to conduct the transit and microlensing exoplanet surveys.
The one of seven 30cm telescopes will be used for the microlensing survey toward the GB.
The ET is planning to measure the masses of FFPs by the space parallax in collaboration with ground base telescopes.
\cite{Ge2022} estimated that ET will detect about 600 FFP events, of which about 150 will have mass measurements.
Our mass function is about a factor of 1.4 higher normalization than that assumed in \cite{Ge2022} with similar slope.
However, they assumed a flat MF for $ \le 1M_\earth$, while we continued the power law slope down to the lower limit
of $0.1 M_\earth$.
This renormalization will update the expected yield of $\sim$840 FFPs with masses down to that of Mars (including $\sim$210 with masses $\le M_\earth$).

\acknowledgments
We thank the anonymous referee for the useful suggestions.
We are grateful to S. Ida, S. A. Johnson and K. Masuda for helpful comments.
The MOA project is supported by JSPS KAKENHI Grant Number JSPS24253004, JSPS26247023, JSPS23340064, JSPS15H00781, JP16H06287, JP17H02871 and JP22H00153.
NK was supported by the JSPS overseas research fellowship.
DPB acknowledges support from NASA grants 80NSSC20K0886 and 80NSSC18K0793.

\appendix

\section{Comparison of Integrated Detection Efficiency to KMT Formula}\label{sec-append}

\begin{figure}
\begin{center}
\includegraphics[scale=0.4,keepaspectratio]{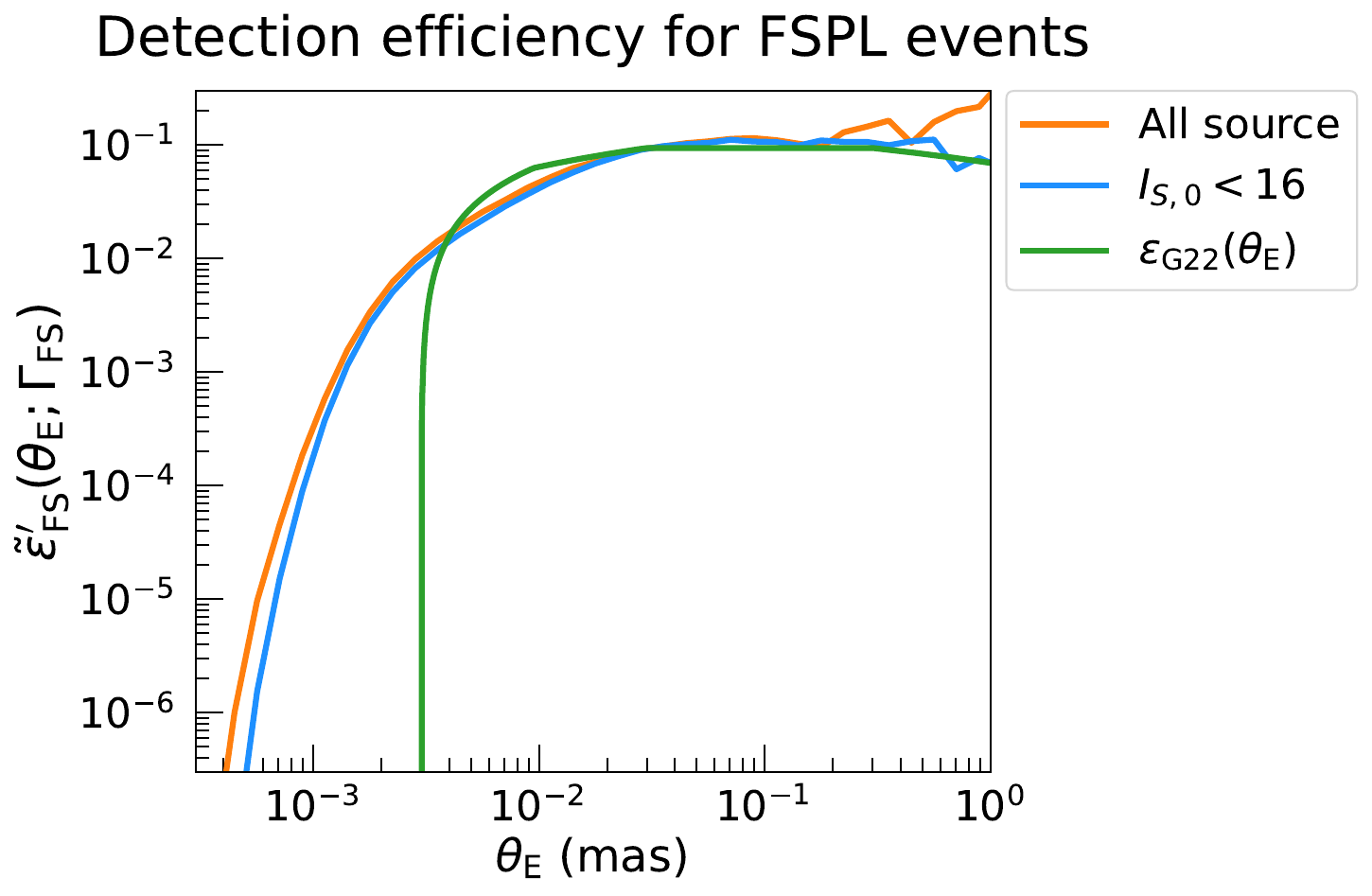}
\caption{
  \label{fig:EFF_thetaE}
Integrated detection efficiencies for events with $u_0  < \rho$ and a significant finite 
source signal as a function of $\theta_{\rm E}$ for all sources (orange line) and 
giant sources with $I_{\rm s,0}<16$ mag (blue line) from MOA \citep{Koshimoto2023}.
and FSPL events from KMTNet (green line) \citep{Gould2022}.  
$I_{\rm s,0}<16$ mag is the limiting magnitude used by \cite{Gould2022},
This is similar to the $\tilde{\epsilon}_{\rm FS} (\theta_{\rm E} ; \Gamma)$ shown in Figure 8 
of \citetalias{Koshimoto2023}, except that \citetalias{Koshimoto2023} do not include the $u_0  < \rho$
condition. These are only for the comparison 
to \cite{Gould2022} and not used for our analysis.
}
\end{center}
\end{figure}

\citetalias{Koshimoto2023} calculated the integrated detection efficiency for our FSPL event sample, 
$\tilde{\epsilon}_{\rm FS} (\theta_{\rm E} ; \Gamma)$. This integrated detection efficiency is similar to the integrated detection efficiency,
$\tilde{\epsilon} (t_{\rm E} ; \Gamma)$, discussed in Sections~\ref{sec:L_longtE} and \ref{sec:L_shorttE}, except that
it has been integrated over $t_{\rm E}$ instead of $\theta_{\rm E}$ for events with a significant finite source
signal, i.e., a measurement of $\rho$.
The ``relative detection efficiency" adopted for KMTNet by \citet{Gould2022} is actually 
a relative integrated detection efficiency in our nomenclature, which we think is more accurate. 
Their relative detection efficiency seems to be a ratio of the number of the events with the detection of FS effect relative to 
the number of events with $u_0 < \rho$ while our $\tilde{\epsilon}_{\rm FS} (\theta_{\rm E} ; \Gamma)$ in \citetalias{Koshimoto2023}  
is that relative to all events with $u_0  \le 1$.
To compare these, we calculated  the integrated detection efficiency with FS effect relative to the events with $u_0  < \rho$, 
denoted as
$\tilde{\epsilon}^\prime_{\rm FS} (\theta_{\rm E} ; \Gamma_{\rm FS})$
and shown in Figure~\ref{fig:EFF_thetaE}.
The integrated detection efficiency depends on the FFP mass function, so  we have used
used our best fit mass function to calculate these curves.
This figure shows the MOA integrated detection efficiency as a function of $\theta_{\rm E}$ 
for all sources (orange) and for 
giant sources with $I_{\rm s,0}<16$ (blue). This is the same limit on $I_{\rm s,0}$ as used by   
KMTNet \citep{Gould2022}, for their analysis of FSPL events. The green curve shows KMTNet's adopted relative integrated
detection efficiency. Both the MOA $I_{\rm s,0}<16$ curve and the KMTNet curves are normalized to match the 
MOA all-source integrated detection efficiency at $\log_{10}(\theta_{\rm E}) = -1.5$.

The MOA sensitivity for giant sources is less than that for all sources at small $\theta_{\rm E}$ because the 
large $\theta_*$ values for giant sources can significantly reduce the peak microlensing magnification.
However, the sensitivity curve for KMTNet is very different from that of MOA, with a much sharper cutoff at
small  $\theta_{\rm E}$. This is partly because they directly cut off their integrated detection efficiency with a
cut excluding events with $\theta_{\rm E}< 3\,\mu$as. 
\cite{Gould2022} describe this cut by saying ``we complete this 
function linearly by imposing a threshold at $\theta_{\rm E} = 3\,\mu$as, which is supported by the fact that all 
four FFPs are pressed up close to this limit\rlap." It is difficult to understand they would need a cut like this given
the sensitivity calculated for our analysis. Similarly, 
two of the four FFP events with finite source effects found by OGLE \citep{Mroz18,Mroz19b,Mroz20b,Mroz20c}
have $\theta_{\rm E}< 3\,\mu$as (see Table~\ref{tbl:candlistFFP}) even though they have source stars with $I_{\rm s,0}<16$.
Perhaps the rationale for this cut that requires $\theta_{\rm E}> 3\,\mu$as is to make their analysis consistent
with their power-law prior assumption of $0.9\lesssim p\lesssim 1.2$. However, if this is the reason
for this cut, it would raise the question as to why KMTNet has not been able to find events with $\theta_{\rm E} < 3\,\mu$as
in contrast to MOA and OGLE who clearly have sensitivity well below this limit with bright sources with $I_{\rm s,0}<16$.
It would be helpful to see a full analysis for the KMTNet data set including a complete detection efficiency analysis
that includes both the $t_{\rm E}$ and $\theta_{\rm E}$ dependence.

Note that our analysis does not use this integrated detection efficiency that depends only on $\theta_{\rm E}$. This
integrated detection efficiency is included only for comparison with the \citet{Gould2022} analysis.
 

\end{document}